\documentclass[10pt,journal,compsoc]{IEEEtran}
\pdfoutput=1
\usepackage{booktabs} % For formal tables
\usepackage{subfigure}
\usepackage{blindtext}
\usepackage{graphicx}
\usepackage{csquotes}
\usepackage{cite}
\usepackage{caption}
\usepackage{amssymb}
\usepackage{amsmath}
\usepackage{siunitx}
\usepackage{microtype}
\usepackage{multicol}
\usepackage{upgreek}
\usepackage{booktabs,tabularx,ragged2e,url}
\usepackage[english]{babel}
\usepackage{fancyhdr}
\usepackage[noend]{algpseudocode}
\usepackage{float}
\usepackage{blindtext}

\usepackage{nomencl}
\makenomenclature

\usepackage[ruled,linesnumbered,commentsnumbered]{algorithm2e}

\SetAlFnt{\small}
\SetAlCapFnt{\small}
\SetAlCapNameFnt{\small}
\SetAlCapHSkip{0pt}
\IncMargin{-\parindent}

\usepackage[hang,flushmargin]{footmisc}

\usepackage{xcolor,colortbl}
\definecolor{Gray}{gray}{0.90}
\newcolumntype{a}{>{\columncolor{Gray}}c}

\usepackage{xcolor,soul}
\definecolor{light-gray}{gray}{0.95}
\sethlcolor{light-gray}

\graphicspath{{Images/}}
\newcommand{\fig}[1]{Fig.~\ref{#1}}
\newcommand{\tb}[1]{Table~\ref{#1}}
\newcommand{\eq}[1]{(\ref{#1})}

\usepackage{enumitem,booktabs}
\usepackage[referable]{threeparttablex}
\renewlist{tablenotes}{enumerate}{1}
\makeatletter
\setlist[tablenotes]{label=\tnote{\alph*},ref=\alph*,itemsep=\z@,topsep=\z@skip,partopsep=\z@skip,parsep=\z@,itemindent=\z@,labelsep=.2em,leftmargin=*,align=left,before={\footnotesize}}
\makeatother

\DeclareCaptionLabelSeparator{periodspace}{.\quad}
\begin{document}

\begin{titlepage}
  \begin{center}
    {\huge \bfseries End-to-End Memristive HTM System for Pattern Recognition and Sequence Prediction  \par
    }

    \bigskip%
    by

    \bigskip%
    \large
    Abdullah M. Zyarah,
        Kevin Gomez, and
        Dhireesha Kudithipudi

    %%%%%
    \end{center}
    \large
     \bigskip\bigskip\bigskip\bigskip%
     \Large
\textcolor{red}{Note: The post-production title of the paper is changed to "Neuromorphic System for Spatial and Temporal Information Processing"}
    \bigskip\bigskip\bigskip\bigskip%

\noindent \textbf{Bibtex citation:} \newline
@ARTICLE \{9109629,\newline
author=\{Zyarah, Abdullah M and Gomez, Kevin and Kudithipudi, Dhireesha\},\newline
journal = \{IEEE Transactions on Computers\},\newline
title = \{Neuromorphic System for Spatial and Temporal Information Processing\},\newline
year = \{2020\},\newline
doi = \{10.1109/TC.2020.3000183\},\newline
publisher = \{IEEE Computer Society\}\newline
\}
  
\bigskip \bigskip% 
\noindent
\textbf{\noindent Plain-text citation:} \newline
A. M. Zyarah, K. Gomez and D. Kudithipudi, "Neuromorphic System for Spatial and Temporal Information Processing," in IEEE Transactions on Computers, doi: 10.1109/TC.2020.3000183.

  \vspace*{\fill}
  \end{titlepage}

\title{End-to-End Memristive HTM System for Pattern Recognition and Sequence Prediction}

% >>> The HTM system capability is not limited to prediction 
% End-to-End Memristive Hierarchical Temporal Memory System for Prediction

\author{Abdullah M. Zyarah,
        Kevin Gomez,
        Dhireesha Kudithipudi,~\IEEEmembership{Senior member,~IEEE}% 
\IEEEcompsocitemizethanks{
\IEEEcompsocthanksitem This work is sponsored in part by Seagate Technology.
\IEEEcompsocthanksitem Abdullah M. Zyarah is with the Neuromorphic AI Lab, Department of Computer Engineering, Rochester Institute of Technology, NY 14623, USA (E-mail: amz6011@rit.edu).
\IEEEcompsocthanksitem Kevin Gomez is with Seagate Research Group, Seagate Technology, MN 55379, USA (Email: kevin.gomez@seagate.com)
\IEEEcompsocthanksitem Dhireesha Kudithipudi is with the Neuromorphic AI Lab, Department of Electrical and Computer Engineering, University of Texas at San Antonio, TX 78249 USA (E-mail: dk@utsa.edu).
}% <-this % stops an unwanted space
}

\IEEEoverridecommandlockouts
\IEEEpubid{\makebox[\columnwidth]{~XXXX-XXXX
\copyright2020
IEEE \hfill} \hspace{\columnsep}\makebox[\columnwidth]{ }} 

% The paper headers
\markboth{IEEE TRANSACTIONS ON COMPUTERS, 2020}%
{Shell \MakeLowercase{\textit{et al.}}: Bare Demo of IEEEtran.cls for Computer Society Journals}
%#############################################################################

\IEEEtitleabstractindextext{%
\begin{abstract}
Neuromorphic systems that learn and predict from streaming inputs hold significant promise in pervasive edge computing and its applications. In this paper, a neuromorphic system that processes spatio-temporal information on the edge is proposed. Algorithmically, the system is based on hierarchical temporal memory that inherently offers online learning, resiliency, and fault tolerance. Architecturally, it is a full custom mixed-signal design with an underlying digital communication scheme and analog computational modules. Therefore, the proposed system features reconfigurability, real-time processing, low power consumption, and low-latency processing. The proposed architecture is benchmarked to predict on real-world streaming data. The network's mean absolute percentage error on the mixed-signal system is 1.129X lower compared to its baseline algorithm model. This reduction can be attributed to device non-idealities and probabilistic formation of synaptic connections. We demonstrate that the combined effect of Hebbian learning and network sparsity also plays a major role in extending the overall network lifespan. We also illustrate that the system offers 3.46X reduction in latency and 77.02X reduction in power consumption when compared to a custom CMOS digital design implemented at the same technology node. By employing specific low power techniques, such as clock gating, we observe 161.37X reduction in power consumption.
\end{abstract}

}

% Suggested alternative titles: 
% (Towards) Neuro-inspired computation processor for spatial and temporal information processing
% Neuro-inspired edge-computing processor for spatial and temporal information processing
% HTM-Based Neuro-Computational Mixed-Signal Processor for Spatial and Temporal Information processing 
% Mixed-Signal HTM-Based Neuromorphic System for Spatial and Temporal Information processing 

% Names: Neorphic, 

% make the title area
\maketitle
\IEEEdisplaynontitleabstractindextext
% \IEEEdisplaynontitleabstractindextext has no effect when using
% compsoc or transmag under a non-conference mode.
\IEEEpeerreviewmaketitle

%#####################################################

\IEEEraisesectionheading{\section{Introduction}\label{sec:introduction}}
\IEEEPARstart{O}{ver} the course of the last decade, there has been a profound shift in artificial intelligence (AI) research, where biologically inspired computing systems are being actively studied to address the demand for energy-efficient intelligent devices. Biologically inspired systems, such as hierarchical temporal memory (HTM)~\cite{hawkins2005intelligence, hawkins2009sequence}, have demonstrated strong capability in processing spatial and temporal information with a high degree of plasticity while learning models of the world. HTM also exhibits natural compatibility for continuous online learning~\cite{george2009towards}, noise and fault tolerance~\cite{hawkins2016neurons}, and low power consumption achieved through sparse neuronal activity~\cite{padilla2015analysis, zyarah2019neuromemrisitive}. These properties make the algorithm attractive for a wide range of applications such as visual object recognition and classification~\cite{HTM_Zeta1_HW_7,xing2012bio}, prediction of data streams~\cite{el2019performance}, natural language processing and anomaly detection~\cite{lavin2015evaluating}. Despite the fact that HTM is an attractive algorithm, it demands high computational power that cannot be fulfilled by conventional von Neumann architectures. This is because the innate HTM architecture, which is composed of thousands of neuronal circuits, requires a high-level parallelism in information processing. One may map the HTM algorithm onto a GPU. A GPU can provide the necessary parallelism, but it fails to provide satisfactory performance and demands a large power budget~\cite{li2019design}. To this end, several research groups have attempted to develop specialized custom hardware designs to run the HTM algorithm efficiently and affordably~\cite{krestinskaya2018hierarchical}. While some of the previous designs focused only on the spatial aspect of the HTM~\cite{streat2016non, kerner2017hierarchical, ibrayev2016design}, other endeavors incorporated both the spatial and temporal models in the same design. For instance, in 2015, Zyarah et al. implemented the HTM algorithm including the spatial and temporal aspects~\cite{zyarah2015design}. The implemented network incorporates 100 mini-columns with 3 cells each, and is verified for image classification and sequence prediction. Furthermore, it supports synthetic synapses, which are realized with distributed memory blocks, to enable synaptic pathway dynamics. The authors also optimized their design further in~\cite{zyarah2019neuromorphic}. Weifu Li et al.~\cite{li2016hardware}, proposed a full architecture of the HTM algorithm in 2016. The proposed architecture is composed of 400 mini-columns (2 cells in each mini-column) connected in point-to-point format to the HTM input space, which eventually causes the mini-column to be in an active mode even when there is insignificant activity (noise) in the input space. When it comes to HTM memristor-based analog and mixed-signal designs, in 2016, Fan et al. implemented the first generation of HTM, HTM-Zeta. The authors proposed RCN (resistive-crossbar networks) pattern matching modules with core processing unit named, spin-neurons~\cite{fan2015hierarchical}. The network operation is verified for image classification in an offline fashion as the proposed design does not support online learning. In 2018, Krestinskaya et al. presented the full analog design of the HTM, but the temporal aspect of the implementation does not match that in the HTM sequence memory as it depends on the class map concept which matches the stored patterns with the test ones (unseen input samples)~\cite{krestinskaya2017hierarchical}.
% The network may also be incapable of capturing temporal information 
To the best of our knowledge, there is no full custom mixed-signal design of the HTM algorithm in literature with underlying digital communication scheme and analog computational modules. Such a design should include the necessary reconfigurability, low energy-delay product, and a robust communication scheme, in one platform. It is important to mention here that such architectures have been explored in the context of spiking neural networks (SNNs)~\cite{benjamin2014neurogrid}, where the communication scheme is realized with address event representation (AER), developed by Mahowald in 1992~\cite{mahowald1992vlsi}. AER takes advantage of sparse neuronal activity and high-bandwidth VLSI to enable time-multiplexed communication. Hence, it reduces the number of connections between sending and receiving neuronal arrays from $n$ to $\log_2 n $~\cite{vogelstein2003spike}. It turns out that AER is considered as an effective approach for point-to-point connections, but not for complex networks with sparse connections. The complex network connectivity is solved through the enhanced AER proposed by Goldberg et al.\cite{Goldberg2001}. The enhanced AER uses look-up tables (LUTs) to describe the connectivity network between two sets of neuronal arrays. The LUT contains the sender address, destination address, and the probability of connectivity. Thus, complex networks even sparse ones can easily be implemented. However, the enhanced AER demands a large amount of memory and this makes it unsuitable for power and area constrained devices. 
% increasing the number of senders/receivers leads to an increase in the memory size which eventually leads to more area and power consumption. 
% - The memory size required to described enhanced AER = Number of senders x Number of receivers. In HTM with 4096 cells, this cost = 16M-Z of memory. Z is the memory register size
Therefore, this paper also proposes a synthetic synapses representation (SSA) communication scheme, which leverages the linear feedback shift registers (LFSR)s to describe the sparse connections among neurons. Using the LFSRs eliminates the need for memory-based address description as the addresses between neurons are generated rather than stored. This results in a considerable reduction in the network area and power consumption. 

Specific key contributions of this paper are as follows:
\begin{itemize}
    \item Developing a memristor-based mixed-signal neuromorphic system of the HTM network including both the spatial pooler and temporal memory. 
    \item Synthetic synapses representation (SSR) communication scheme is proposed to virtually formulate and prune the physical synaptic connections in the HTM network. 
    \item System-level analysis of the performance, lifespan, area, and power consumption with respect to a CMOS only implementation is performed. 
\end{itemize}

% The remainder of this paper is organized as follows: Section 2 provides a brief background about the HTM algorithm. The system design of the HTM including the core units is presented in Section 3. Section 4 explains the synthetic synaptic representation and its advantage over the address event representation. Section 5 and 6 discuss the experimental methodology and the achieved results, respectively. The paper conclusions are presented in section 7.

%%%%%%%%%%%%%%%%%%%%%%%%%%%%%%%%%%%%%%%%%%%%%%%%%%%%%%%%%%%%%%%%%%%
\section{Hierarchical Temporal memory}
HTM is a biomimetic algorithm that aims to develop a computational framework capturing the structure and the algorithmic properties of the human neocortex. Structurally, the algorithm is composed of hierarchical ascending layers of cellular regions that enable the network to capture spatial and temporal information, shown in~\fig{htm_hierar}. Each region in the HTM is composed of building blocks, namely cells, which are arranged in columns to model biological mini-columns. The cell in HTM is just an abstract model of the excitatory pyramidal neurons. As pyramidal neurons, each cell has hundreds of synaptic connections grouped into three integration zones (or segments): proximal, distal, and apical\footnote{A cell in HTM typically has one proximal segment (shared with other cells of the same mini-column) and multiple distal and apical segments.}~\cite{hawkins2016neurons, cui2016comparative}. The proximal segment
%\footnote{In HTM, the cells within one mini-column share the same proximal segment.} 
is dedicated to receive feed-forward input i.e. observe the cellular activities in the lower layers in the hierarchy, or sensory input. Typically, activities detected on proximal segments leads to generation of neuronal action potential. The distal and apical segments, on the other hand, are dedicated to observe the cellular activities of the neighboring cells in the same region (contextual input) and higher levels in the hierarchy (feedback input), respectively. Unlike the proximal segment, the cellular activities detected by distal and apical segments lead to NMDA spikes~\cite{cui2016continuous}. The NMDA spikes slightly depolarize the cell without generating an action potential, giving the cell a competitive advantage in responding to future input representations~\cite{cui2017htm}.

\fig{htm_hierar} shows a high-level diagram of the HTM network equipped with a data encoder and multiple classifiers. The encoder transforms sensory information into binary representations, while the classifiers map the HTM output to the corresponding class labels (SDR classifier) and identify anomalies (anomaly classifier). The mixed-signal design of the SDR classifier has been developed in our previous work~\cite{zyarah2019neuromemrisitive}. Thus, this work will emphasize the design and implementation of a single HTM region\footnote{The hierarchical structure of the HTM network has not been thoroughly studied yet.}, which is equivalent to realizing the primary sensory region in the supra-granular layers of the neocortex.
Given an HTM region, there are two core operations which capture the spatial and temporal information of a given input, namely the spatial pooler and temporal memory, which are discussed in the following subsections:  

\begin{figure} [h!tb]
\begin{center}
\includegraphics[width = 0.42 \textwidth]{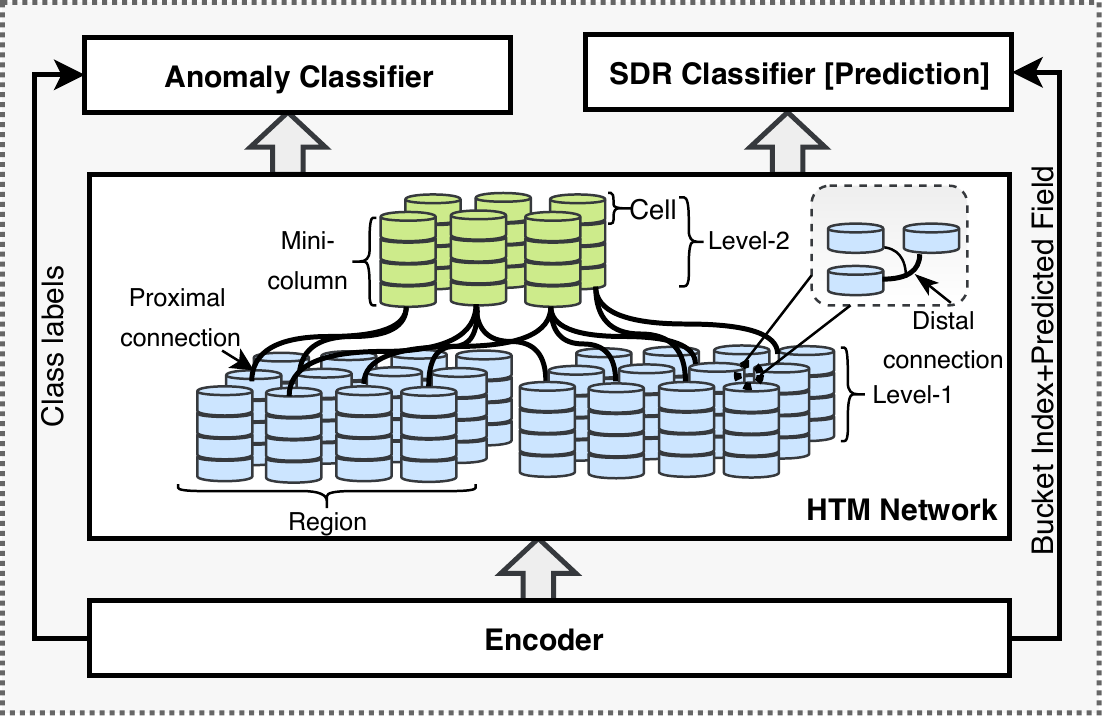}
\caption{High-level architecture of the HTM system with three core units: data encoder, HTM network, and classifiers. The encoder transforms the input data into binary representations. The HTM algorithm learns spatial information and captures temporal transitions, while the classifiers map the HTM output to the corresponding class labels and identify anomalies.\vspace{-3mm}}
\label{htm_hierar}
\end{center}
\end{figure}

% use the term Spatial pooler, instead of SP.
%> I replaced SP with spatial pooler and the same is applied for TM.
\subsection{Spatial Pooler}
The spatial pooler in the HTM is responsible for extracting and learning the spatial patterns of the sequential data. Typically, the spatial pooler models an encoded sensory input, generated by the encoder, using a population of active and inactive mini-columns chosen through a combination of competitive Hebbian learning rules and homeostasis~\cite{cui2017htm}. Typically, the number of active mini-columns is limited to (2-4)\% of the total mini-columns in a given HTM region, resulting in so-called sparse distributed representation (SDR). The SDR in HTM defines the underlying data structure and enables the crucial features of the algorithm such as distinguishing the common features between inputs~\cite{SDR_P}, learning sequences, and making simultaneous predictions~\cite{sparse_whitepaper}. However, selecting the active and inactive mini-columns is determined according to the spatial response of the individual mini-columns to an input. Recall, each mini-column in the HTM observes the pattern activities in the input space using a set of proximal connections. Having a reasonable number of active proximal synapses connected to active bits (namely overlap score) initiates an action potential that enables a mini-column to compete with its neighbour for input representation. By using the $k$-winner-take-all ($k$-WTA) computation principle, the top (2-4)\% mini-columns with the highest overlap scores are activated (become winners) and inhibit their neighbors. The output of the spatial pooler is a binary vector, which represents the joint activity of all mini-columns in the HTM region in response to the current input. The spatial pooler operation can be divided into three distinct phases: initialization, overlap and inhibition, and learning, discussed in our previous work\cite{zyarah2019neuromemrisitive} and briefly described below. 

In the initialization phase (Algorithm~\ref{alg:one}, lines 2-5), which occurs only once, the mini-columns' connections to the input space, synapses' permanences, and boosting factors, are initialized. Let $S_p$ be an $n_c\times n_x$ array holding the proximal synaptic connections between $n_c$ mini-columns and $n_x-$dimensional input space. Similarly, let $\rho_p$ be an $n_c\times n_x$ array that defines the permanence of corresponding potential synapses in $S_p$. Given the $j^{th}$ mini-column, its maximum number of potential synapses ($n_{sp}$) is defined by the non-zero elements in $\vec{s_p}$ ($\vec{s_p}$ is a row vector in $S_p$) whose indices are generated by a pseudo-random number generator, and their permanence values are uniformly initialized at random between `0' and `1'. Initializing the synapses is followed by setting the boosting factor of the individual mini-columns to `1'. After the initialization, the overlap and inhibition phase (lines 8-11) starts in which the feed-forward input is collectively represented by a subset of active mini-columns (winning mini-columns). Selecting the active mini-columns is done after counting mini-columns' active synapses that are associated with active bits in the input space, i.e. overlap scores ($\alpha$). Mathematically, this is achieved by performing a dot product operation between the feed-forward input vector ($\vec{x}^t$) at time $t$ and the active synapses array, as in line 9, where the active synapses array is the result of an element-wise multiplication (denoted as $\odot$) between $S_p$ and $\bar{\rho}_p$. $\vec{b}^t$, here, denotes the boosting factor that regulates mini-column activities.~$\bar{\rho}_p$ is a permanence binary array to indicate the status of each potential synapse, where `1' indicates a connected synapse and `0' an unconnected synapse. Upon the completion of computing the overlap scores, each mini-column overlap score gets evaluated by comparing it to a threshold, $\alpha_{th}$ (line 10). The resulting vector ($\vec{e\alpha}^t$) is an indicator vector representing the nominated mini-columns with high overlap scores. Given an inhibition radius defined by $\xi$, based on the mini-column overlap scores and desired level of sparsity ($\eta$), $n_w$ number of mini-columns will be selected to represent the input, as shown in line 11. After determining the winning mini-columns in $\vec{\Lambda}^t$, the learning phase (lines 13-16) starts to update the permanence values of the winning mini-columns' synapses. The synapses' permanences are updated according to Hebbian rule~\cite{hebb1949organization}. The rule implies that the synapses connected to active bits must be strengthened, increasing their permanence by $P^+_p$, while those connected to inactive bits will be weakened, decreasing their permanence by $P_p^-$, as in line 14, where $\Delta\rho_{p}$ is the change in the permanence array for all mini-columns given an input $\vec{x}^t$, and $\lambda$ denotes the sum of $P_p^+$ and $P_p^-$. After adjusting the synapse's permanence, the boosting factor of each mini-column is updated according to the mini-column's time-averaged activity level ($\bar{a}^t$) and its activity level with respect to its neighbor ($<$$\bar{a}^t$$>$)~\cite{cui2017htm}.
% nx: input space size
% ns: number of synapses per columns
% nc: number of columns in a region
% nw: is the number of winning columns to encode each input sample

\setlength{\textfloatsep}{3mm}
\IncMargin{1em}
\begin{algorithm}[t]
\caption{HTM-Spatial Pooling}
\label{alg:one}
\KwIn{$\vec{x}^t \in \mathbb{R}^{n_x}_{\{0,1\}}$, where $\vec{x}^t \subset X^t$ and $X^t \in \mathbb{R}^{n_x \times n_n}_{\{0,1\}};$
}
\KwOut{$\vec{\Lambda}^t \in \mathbb{R}^{n_c}_{\{0,1\}}$\scriptsize{\tcc*[r]{$n_c$:Number of mini-columns}}} 
// Initialization:~~~~~~~~~~~~~\tcc*[h]{\scriptsize{$n_x$:Input vector length}}\\
$S_{ind} \sim$ rand.pseudo, where $S_{ind} \in \mathbb{N}^{n_c\times n_{sp}}_{\{1, n_x\}}$ \;
$S_p[S_{ind}] \leftarrow 1, \text{where}~S~\text{and}~\rho \in  \mathbb{R}^{n_c \times n_x}$ \;
$\rho_p[S_{ind}] \sim$ rand.uniform[0,1]\;
$\vec{b}^t \in \mathbb{R}^{n_c}$, where $\forall~b^t[j] = 1$\;
\Repeat{$t > n_n$}
	{
	// Overlap and Inhibition:\\
 	$\bar{\rho}_p \leftarrow \text{I}(\rho_p \geq P_{th})$ \;				
 	$\vec{\alpha}^t \leftarrow \vec{b}^t \odot \big[(S_p \odot \bar{\rho}_p) \cdot \vec{x}^t \big]$ \;
 	$\vec{e\alpha}^t \leftarrow \text{I}(\vec{\alpha}^t \geq \alpha_{th})$ \;
 	$\vec{\Lambda}^t \leftarrow kmax(\vec{e\alpha}^t, \eta, \xi)$;~~~~~\tcc*[h]{\scriptsize{$kmax$:$k$-WTA function}} \\
 	// Learning:\\
 	\If{Learning == 'Enable'} 
 	{
  		{$\Delta\rho_p \leftarrow \vec{\Lambda}^t.transpose \odot (S_p \odot \bar{\rho}_p) \odot (\lambda\vec{x}^t - P_p^-$)\;
  		$\vec{b}^t \leftarrow e^{-\gamma(\bar{a}^t - <a^t>)}$ \;}
 		}
	}
\end{algorithm}
\DecMargin{1em}

\subsection{Temporal Memory}
The temporal memory in the HTM is mainly dedicated to learn time-based sequences and to make predictions. The temporal memory operates at the cells level, specifically, the cells of the winning mini-columns. When a mini-column becomes active, at least one of its cells is selected to be active to represent the input contextually. This usually depends on whether the cells within the winning  mini-columns are predicting the incoming input. If a winning mini-column has a predictive cell, that cell becomes active and inhibits other cells within the same mini-columns from being active. Otherwise, the joint activation of all cells within the mini-column represents the input and this is known as massive neurons firing or bursting. However, once a cell is activated, it forms lateral connections with the cells that were active in the previous time step. Patterns recognized by lateral connections lead to a slight depolarization of the cell soma (predictive state), subsequently predicting the upcoming events. Typically, the lateral connections are grouped into distal segments. A cell in HTM can have more than one distal segment and this grants the cells the capability to predict more unique temporal patterns. The operation of the temporal memory can be divided into three phases: mini-columns evaluation, prediction, and learning phase, described in Algorithm~\ref{alg:two}\footnote{Forming and pruning lateral connections are not discussed in the algorithm to avoid complexity.}.

During the mini-columns evaluation phase (Algorithm~\ref{alg:two}, Line 4-15), the active cells within the winning mini-columns are selected to represent the input within its context. Let $n_m$ be the number of cells in each mini-column, and $A^t \in R^{n_m \times n_c}_{\{0,1\}}$ is a binary array that represents the region cells' activity, where `1' indicates an active cell and `0' is inactive. Similarly, let $\pi^t$ be also a binary array that has the same size of $A$, and the active bits in $\pi$ refers to the predictive cells. An $i^{th}$ cell within the $j^{th}$ mini-column is set to be active if $\vec{\Lambda}_j^t=1$ and the cell was in the predictive state in the previous time step i.e. $\pi^{t-1}_{ij} =1$. Otherwise, bursting (all cells within the $j^{th}$ mini-column are set to be active) will take place. 

In the second phase of temporal memory, prediction (line 17-35), the status of the cells for the next time step is evaluated. This is done via observing the distal segments activation level ($\alpha$). Let $D_{ij}$ represent a group of distal segments that belong to the $i^{th}$ cell within the $j^{th}$ mini-column, where a segment in $D_{ij}$ indexed by $d$. If $\bar{\rho}_{ij}^d$ is the active distal synaptic connection within the $d^{th}$ segment, and $\bar{S}_{ij}^{d}$ holds its distal connections that are connected to active bits in $A^t$, the $d^{th}$ distal segment is set to be active segment if its $||\bar{\rho}_{ij}^{d} \cdot \bar{S}_{ij}^{d}||_1$ is greater than the segments activation threshold, $D_{th}$. Otherwise, the segment is set to a matching state if it has at least one synapse connected to an active cell in $A^t$. Once the status of the distal segments are determined, the cells with active distal segments are set to be in the predictive state. It is important to mention here that occasionally cells in HTM may incorrectly predict patterns. In such scenario, these cells need to have their synaptic strength reduced to lower the likelihood of incorrect prediction (as in lines 19-23). After evaluating the cells' segments, their synaptic connections are updated, which occurs during the learning phase (lines 38-47).

As aforementioned, the learning in HTM follows Hebbian's rule and it is applied solely to active cells. Given $a^t_{ij} \in A^t$, where $a^t_{ij}=1$ and has an active segment, then all the synaptic connections that are laterally connected to previous active cells are potentiated, while those that are connected to inactive cells are depressed. This implies that the permanences of the distal synaptic connections, $\rho^d_{ij}$, are increased by $P^+$ when they are connected to active cells, otherwise, they are decreased by $P^-$.

\IncMargin{1em}
\begin{algorithm}[t]
\caption{HTM-Temporal Memory}
\label{alg:two}
\KwIn{$\vec{\Lambda}^t \in\mathbb{R}^{n_c}_{\{0,1\}}$\scriptsize{\tcc*[r]{$n_c$: Number of columns}}}
\KwOut{$A^t \in \mathbb{R}^{n_m\times n_c}_{\{0,1\}}$ \scriptsize{\tcc*[r]{$n_m$: Number of cells}}} 
$zeros\_cnt$ = 0\;
\Repeat{$t > n_n$}
{
\# Phase-1: Mini-columns evaluation: \\
\For{$j \leftarrow 1$ \KwTo $n_c$}
{
    \uIf{$\vec{\Lambda}^t[j] == 1$}
    {
        \For{$i \leftarrow 1$ \KwTo $n_m$}
        {   
            \uIf{${\pi}^{t-1}[i,j] == 1$}
            {
                $A^t[i,j] \leftarrow 1$\;
            }
            \uElse
            {
                $zeros\_cnt \leftarrow zeros\_cnt+1$\;
            }
        }
    \uIf{$zeros\_cnt == n_m$}
    {
        $A^t[i,j] \leftarrow 1, \forall i$\;
    }
    $zeros\_cnt$ = 0\;
    }
}
\noindent \# Phase-2: Prediction:\\
\For{$j \leftarrow 1$ \KwTo $n_c$}
{
    \For{$i \leftarrow 1$ \KwTo $n_m$}
    {
    \uIf{$\vec{\Lambda}^t[j] == 0$ and $\pi^{t-1}[i,j] == 1$}
        {
            \For{$d \leftarrow 1$ \KwTo $n_d$}
            {    
                \uIf{$D[i,j][d].MatchingSegment$}
                    {
                        $\Delta \rho[i,j][d] \leftarrow (A^{t-1} \odot S[i,j][d]) \times \frac{P^+}{10}$\;
                    }
                }
            }
    \uElseIf{$\vec{\Lambda}^t[j] == 0$}  {
        
            \For{{$d \leftarrow 1$ \KwTo $n_d$}}
            {
            $\bar{\rho}[i,j][d] \leftarrow \text{I}(\rho[i,j][d] \geq P_{th})$ \;
            $\bar{S}[i,j][d] \leftarrow A^t \odot S[i,j][d]$ \;
            $\alpha^t \leftarrow ||\bar{S}[i,j][d] \cdot \bar{\rho}[i,j][d]||_1$ \;
            \uIf{$\alpha^t \geq D_{th}$}
            {
                $D[i,j][d].ActiveSegment \leftarrow 1 $\;
                $\pi^t[i,j] \leftarrow 1$  \;
            }
            \uElseIf{$||A^t \cdot S_{ij}^{d}||_1 > 0$}
            {
                $D[i,j][d].MatchingSegment \leftarrow 1$ \;
            }
            }
        }
    }
}
\# Phase-3: Learning:\\
\For{$j \leftarrow 1$ \KwTo $n_c$}
{
    \uIf{$\vec{\Lambda}^t[j] == 1$}
    {
        \For{$i \leftarrow 1$ \KwTo $n_m$}
        {
            \uIf{$A^t[i,j] == 1$}
            {
                \For{$d \leftarrow 1$ \KwTo $n_d$}
                {
                    \uIf{$D[i,j][d].ActiveSegment == 1$}
                    {
                        $\Delta \rho[i,j][d] \leftarrow \lambda (A^{t-1} \odot S[i,j][d])  - P^-$\;
                    }
                }
            }
        }
    }
}
}
\end{algorithm}
\DecMargin{1em}

%&&&&&&&&&&&&&&&&&&&&&&&&&&&&&&&&&&&&&&&&&&&&&&&&&&&&&&&&&
\section{System Design and Methodology}
\begin{figure}[t]
\begin{center}
\includegraphics[width = 0.48 \textwidth]{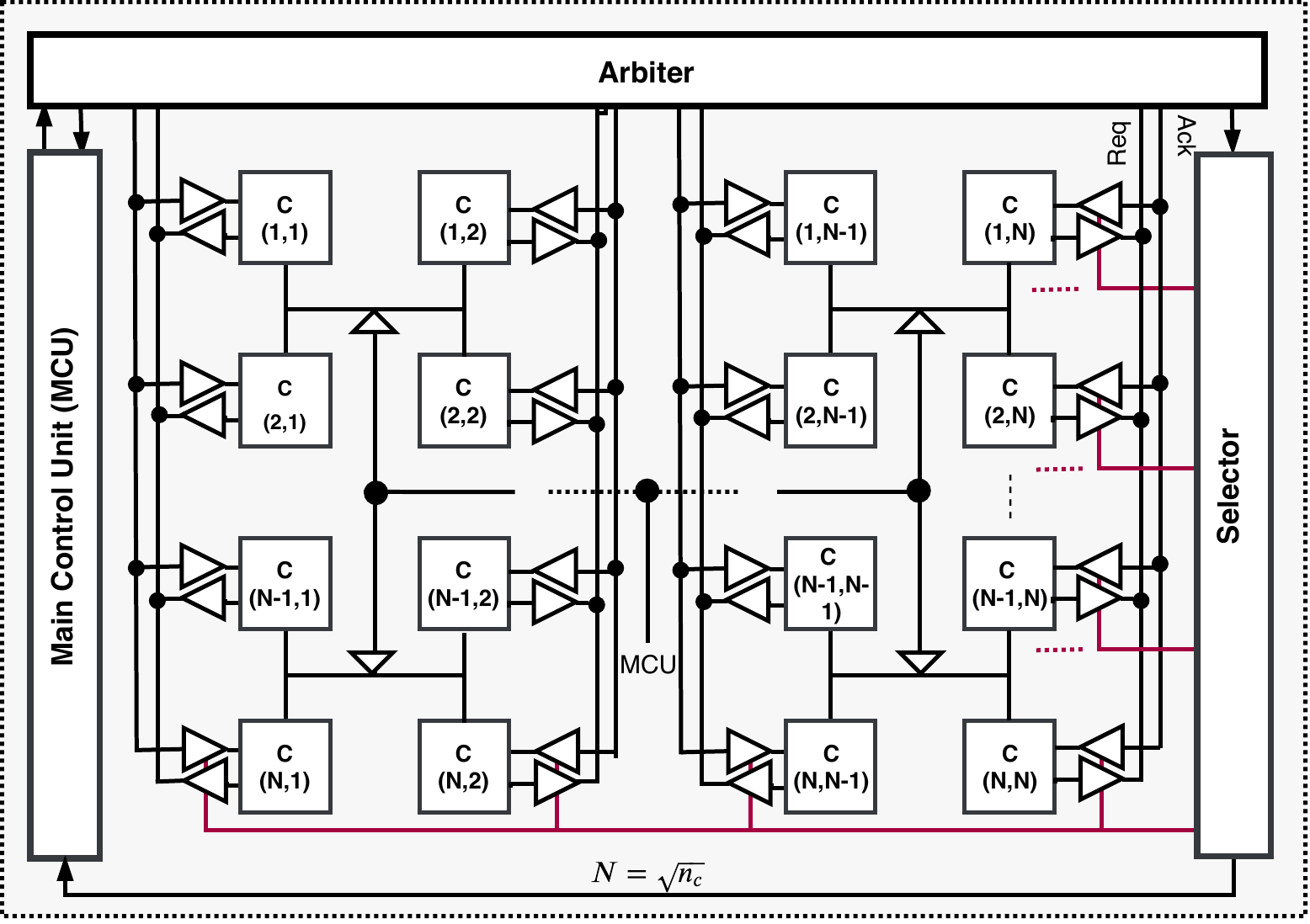}
\caption{High-level architecture of the HTM network, including HTM region with $\sqrt{n_c}\times \sqrt{n_c}$ mini-columns with $n_m$ cells each, a main control unit (MCU), and an arbiter and selector.}
\label{htm_soc}
\end{center}
\end{figure}

\fig{htm_soc} demonstrates the high-level architecture of the developed HTM network\footnote{The feasibility of the HTM network scaling (beyond 1024 mini-columns) can be made possible by adopting the slicing approach proposed in our previous work~\cite{zyarah2019neuromorphic}.} including the core units of the SSR communication scheme. Essentially, there are $\sqrt{n_c} \times \sqrt{n_c}$\footnote{The number of mini-columns assumed in this work is always a power of two, $2^k$, where $k$ is an integer number.} mini-columns with $n_m$ cells each to constitute the HTM region. 
Unlike the mathematical description of the HTM which assumed 2D representation of the region for simplicity, in the hardware design, we consider a 3D architecture of the region to cut down the resources and to simplify the communication scheme considerably. The HTM region is integrated to a main control unit (MCU), and an arbiter and selector. The MCU is dedicated to control data flow, to generate the necessary control signals, and to bridge the region to the input data encoder or other regions in the hierarchy
%\footnote{Using MCU may form a bottleneck that limits the number of mini-columns that can be incorporated in HTM region. However, this bottleneck can be overcome by using the slicing approach, proposed in our previous work in~\cite{zyarah2019neuromorphic}.},
while the arbiter and selector are responsible for regulating data sharing among cells within the region. Here, the interaction among cells is based on the SSR as the cells' activity is sparse in nature, approximately 4.2\%. At a high level, the system works as follows: when the MCU establishes a connection with the data encoder which is done through a hand-shake protocol, it commences receiving the encoded packets. The received packets are routed through the H-Tree to all the region' mini-columns. Here, we used the H-Tree structure to reduce the parasitic capacitance and to minimize the power consumption~\cite{asgari2004low} of the developed system. However, there are two H-Trees, one is a digital bus (34-bit width, $1+\log_2 n$ lines are used by the cells, where $n = n_c \times n_m$) driven by the MCU and the cells to share data. The other one (not shown in \fig{htm_soc}) is an analog line to enable mini-columns to compete against each other for input representation. When the winning mini-columns and then the cells are selected, the arbiter and selector are used to broadcast information about the current/previous active cells and their locations in the region so that lateral connections are formed and future predictions are made. In the following subsections, more details about each core unit of the HTM network are provided, while the communication scheme is discussed in a separate section.

\subsection{HTM Mini-Column}
\begin{figure*} [h!tb]
\begin{center}
\includegraphics[width = 0.9 \textwidth]{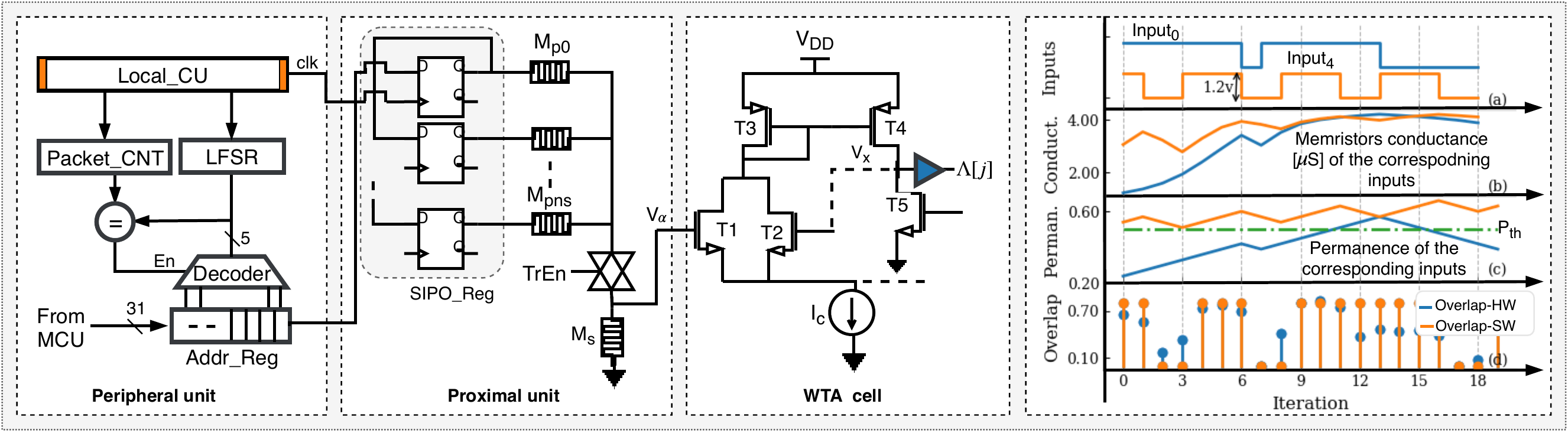}
\caption{(left) The circuit diagram of the HTM mini-column~\mbox{\cite{zyarah2019neuromemrisitive}}. The circuit is composed of a peripheral unit to generate the proximal synaptic connections, a proximal unit to hold the permanences of the synaptic connections, and a WTA cell to enable mini-columns to compete with each other for input representation. (right) The impact of the synaptic permanence (denoted as Perman. for HTM-SW) modulation on the mini-column overlap score as the proximal synapses receive feed-forward input. \vspace{-3mm}}
\label{mini_column}
\end{center}
\end{figure*}

The mini-columns in HTM are responsible for capturing spatial patterns of the feedforward inputs. The HTM mini-column circuit, developed in our previous work~\cite{zyarah2019neuromemrisitive}, is depicted in~\fig{mini_column}-(left). Succinctly, the circuit comprises a peripheral unit, a proximal unit, and WTA cell. In the peripheral unit, the proximal connections are generated and connected to the input space. The proximal unit and the WTA cell hold the proximal connections' permanences and a contesting unit that enables each mini-column to compete with its neighbors for the input representation, respectively. In this work, the input to the mini-column is generated by the HTM random scalar encoder~\cite{purdy2016encoding}, which  encodes every scalar value of the time-series data into a high-dimensional binary vector sorted into small 31-bit packets. This is to minimize data movement and the required storage units. Sequentially, each packet is fetched to the mini-columns and stored into $Addr\_Reg$. When the input packet is stored in the $Addr\_Reg$ and the LFSR generates an address for a location in the received packet, a matching score is stored in the synapses' registers which are modeled by $n_{sp}\times1$ serial-in-parallel-out shift register. Once all inputs are received, the outputs of the synapses' registers are presented to the memristive crossbar word-line where the proximal synapse permanences are stored. The input voltages to the crossbar will be converted into current through the memristor and the output is collected at the crossbar bit-line. The output of the crossbar, which modulates the mini-column overlap score to current, is then boosted. Boosting is done via the use of a sense memristor (M$_s$). However, upon the completion of computing the overlap score ($V_{\alpha j} \equiv \alpha_j$), its value, which is sampled by the sense memristor, is then presented to a WTA circuit (detailed description of the WTA is provided in~\cite{zyarah2019neuromemrisitive}). The WTA performs a $kmax$ operation on $V_{\alpha j}, \forall j$ followed by a thresholding, to generate the final $j^{th}$ mini-column output, ($\Lambda_j$), as given in~\eq{overlap_score} and \eq{col_out}:
\begin{equation}
V_{\alpha_j} = \frac{\sum\limits^{n_s}_{i=1} g_i~V_i }{g_s + \sum\limits^{n_s}_{i=1} g_i}
\label{overlap_score}
\end{equation}
\begin{equation}
\Lambda_j = 
\begin{cases}
1 , & ~V_{x_j} > V_{th},~where~V_{x_j} = f(V_{\alpha j}) \\
0,  & Otherwise
\end{cases}
\label{col_out}
\end{equation}
where $V_i$ denotes the $i^{th}$ input voltage, $g_i$ refers to the conductance of the $i^{th}$ memristor, and $g_s$ is the conductance of the sense memristor. However, once the final output of each mini-column is generated, the learning phase of the spatial pooler starts. As alluded to earlier, the learning in HTM follows Hebbian rule~\cite{hebb1949organization}, which is implemented using Ziksa~\cite{zyarah2017ziksa}, as discussed in~\cite{zyarah2019neuromemrisitive}. Then, the mini-columns' status is relayed to their associated cells to start the next phase, temporal memory. Although the cells are encapsulated within the mini-columns and are considered a part of it, for the sake of clarity and simplicity we dealt with them separately.

\fig{mini_column}-(right) demonstrates the process of computing the overlap score and tuning the proximal synaptic connections for a given mini-column while receiving feed-forward input, shown in \fig{mini_column}-(a). Since the mini-column has a large number of proximal connections, for the purpose of demonstration, we randomly picked only two. The changes in proximal connection permanence for both HTM-SW and HTM-HW models are shown in~\fig{mini_column}-(b) and 4-(c), respectively. Here, it can be observed that any changes in the synapse's permanence below the permanence threshold, $P_{th}$, in the HTM-SW model has no impact on the overlap score, unlike the HW model where there is no explicit threshold blocking the memristors from contributing to overlap score value. Furthermore, the change in the HTM-HW model synaptic permanence (memristors' conductances) tends to be non-linear as compared to the HTM-SW counterpart. However, selecting a memristor device with high conductance range and switching dynamics as required by the HTM theory made the synapses with high conductance states dominate the changes in the overlap level (see~\fig{mini_column}-(d)). This eventually results in almost analogous overlap score\footnote{The overlap scores for the HTM-HW and HTM-SW models are not reported up to scale for the purpose of comparison.} variation for both the SW and HW models.

%--------------------------------
\subsection{HTM Cell} \label{htm_cell_section}
\begin{figure*} [t]
\begin{center}
\includegraphics[width=165mm, height=80mm]{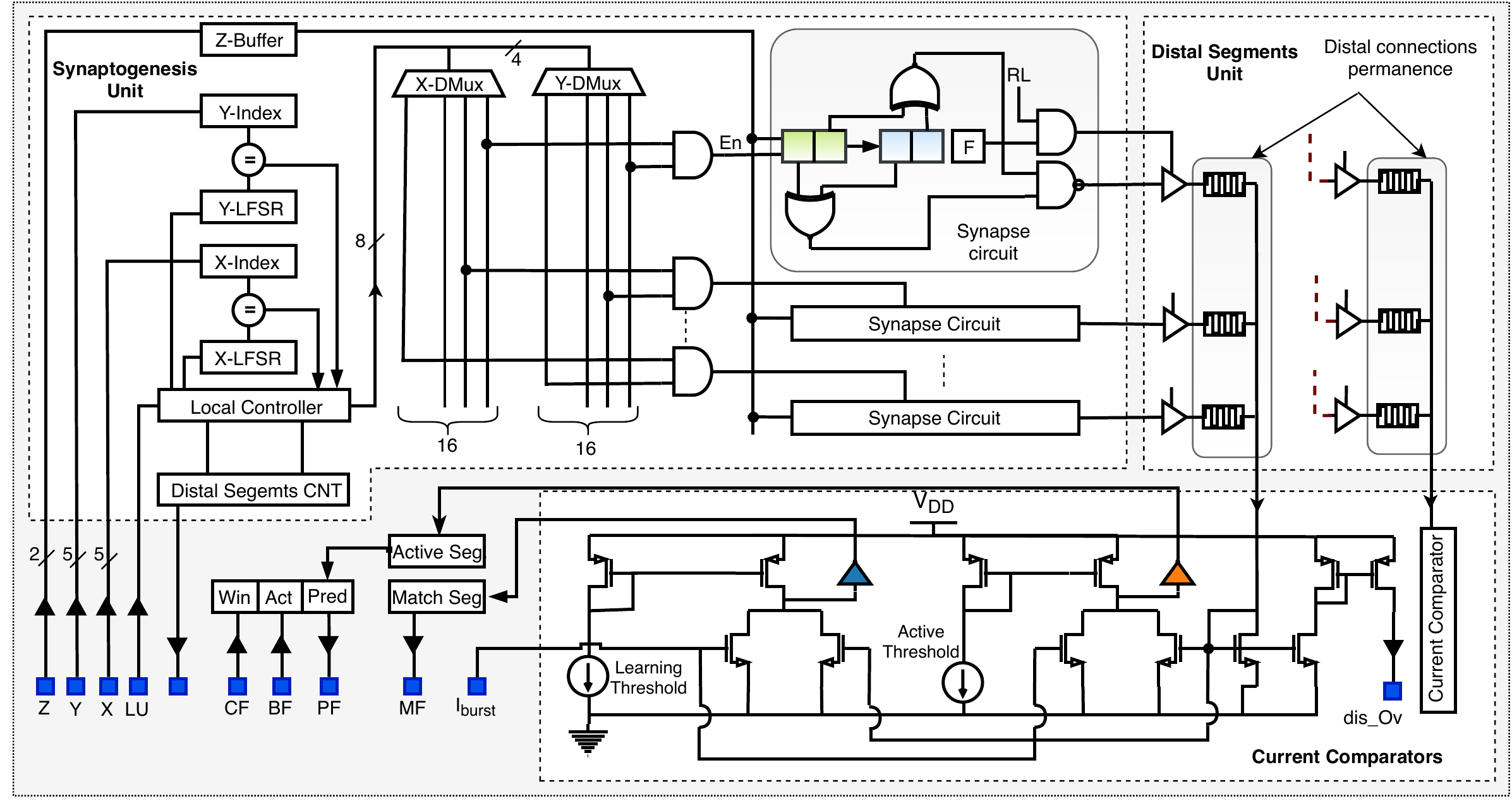}
\caption{The circuit diagram depicting the HTM cell with synaptogenesis unit, which can generate or prune distal segments; a distal dendritic segment to hold the permanence values of the distal connections; and current comparators to evaluate the distal segment activation level and consequently the cell status (predictive or unpredictive).\vspace{0mm}}
\label{htm_cell}
\end{center}
\end{figure*}

The cells in HTM enable the network to capture the temporal patterns, modeling the input representations within their context, and predicting the upcoming events. The HTM cell circuit developed in this work is composed of synaptogenesis unit\footnote{One may share the synaptogenesis unit between multiple cells of the same mini-column to cut-down resources and reduce power consumption, but at the expense of increasing the latency.}, distal segments unit, and current comparators, shown in~\fig{htm_cell}. The synaptogenesis unit is responsible for forming and pruning distal synaptic pathways with the previous active cells. The distal segments unit possesses the permanence values which describe the growth level of the individual distal synaptic pathways, while the current comparators are utilized to evaluate the segments activation level and to determine their states (active or matching) accordingly.

Initially, the cells start with no distal synapses. Once the HTM begins processing the incoming patterns, the distal synapses start forming in the synaptogenesis unit. Given an HTM region arranged into a 3D space, where the $x$ and $y$ axes index the mini-columns in the region and the $z$ axis indexes the cells. When the region receives an input, this causes activation of a population of cells within the region, and in this context, it is referred to as $A^t_{3D}$. If $a^t_{xyz} \in A^t_{3D}$, where $a^t_{xyz}$ is an active cell located at $xyz$,  $a^t_{xyz}$ will form connections with the active cells in $A^{t-1}_{3D}$. Let's assume that the number of active cells in $A^{t-1}_{3D}$ is 4.2\% of $n_c$. Then, if $n_c$=961, $\approx$ 40 cells will be active in each time step, assuming no bursting takes place. The active cell at time $t$ establishes connections with the 40 cells that were active at $t-1$ by forming a distal segment. A cell in HTM can have around 10 or more distal segments, and this enables the network to learn the temporal transitions in sequences. Recall that forming and pruning distal connections in hardware platforms requires high interconnect dynamics which are lacking in most of the existing platforms, especially ASIC designs, hence the virtual description of the synapse became a common approach~\cite{mahowald1992vlsi, zyarah2015reconfigurable}. However, describing the synapses virtually, in most cases, demands a high memory usage to store the sender/receiver addresses. For instance, in HTM's context (assuming there are 961 mini-columns in the region with 4 cells each), if we assume that the address of each cell is represented with 12 bits and the distal connection permanence is represented with 16 bits, having 10 segments with 60 distal connections in each cell costs 16.8kb of memory per cell and more than 64.57Mb for the entire network. Lets assume that the addresses and the permanences are stored in a DRAM implemented in 45nm process. If the energy cost per 32 bits of off-memory access takes 640pJ~\cite{han2017efficient}, having 40 active cells at each time step leads to a total energy consumption of 15.36$\mu$J (first-order approximation). Running the system at 8MHz can result in a power consumption of 122.88W just to access the memory, which is a prohibitive amount of power especially for edge devices with limited power budget. 

% Memory per cell: (12+16)x10x60=16,800 bits

% Energy: Access to 10 segments in 40 cells, where each segment has 60 distal connections. Total energy = 640pJ * 40 * 60 * 10 = 15.36uJ

\begin{figure} [h!tb]
\begin{center}
\includegraphics[width=65mm, height=45mm]{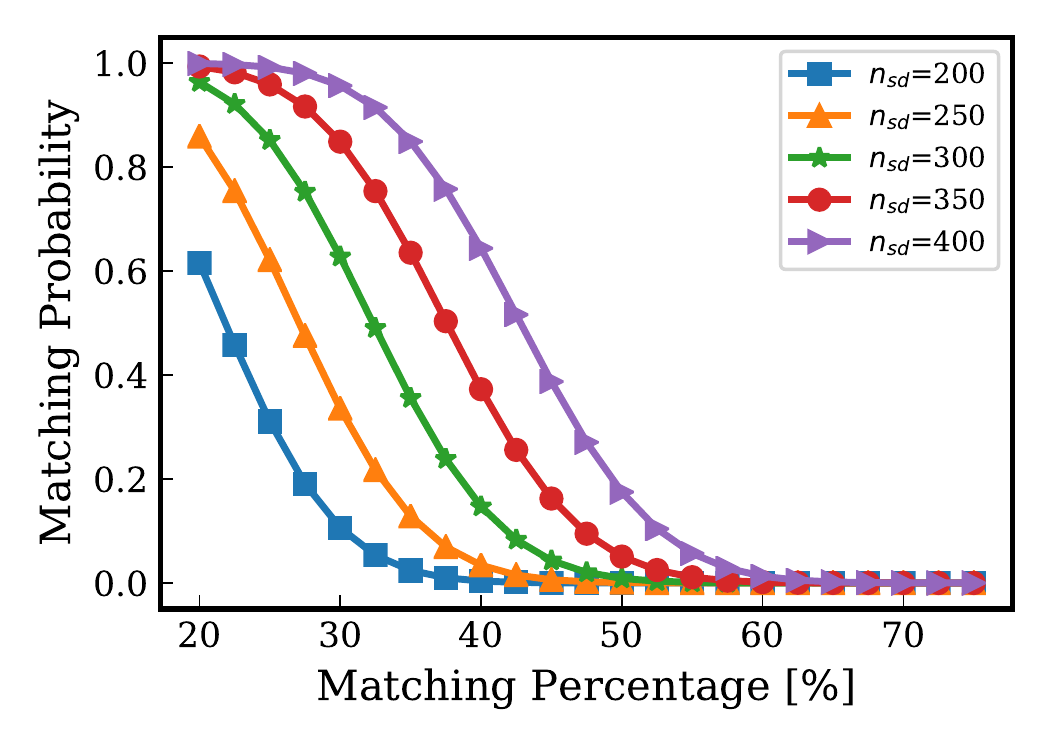}
\caption{The matching probability between a distal segment's addresses generated by LFSRs and the addresses of the active cells in the previous time step for various segment sizes.}
\label{match_prob}
\end{center}
\end{figure}

One possible solution to overcome the above challenge is to reduce the memory usage in each cell. This can be done through modeling the synaptic permanence using analog memristors and leveraging the randomness in forming the distal synaptic connections to generate the addresses rather than storing them. A possible approach to do so is generating the distal segment addresses through the use of LFSRs. To demonstrate this, let's assume that the cell $c_{242}$ is currently active and trying to establish a connection with another cell, $c_{333}$, which was active in the previous time step. The cell $c_{242}$ will receive a packet that holds $c_{333}$ location into 3D space, in this example $333$. Upon receiving the address, the cell, $c_{242}$, begins the matching process in which the cell identifies whether there is a possibility to establish a distal connection with the cell $c_{333}$. The matching process starts by enabling the X-LFSR to generate 16 addresses within one clock cycle\footnote{The cells' LFSRs are clocked with 128MHz, while the system clock is 8MHz.}. The same is applied for the Y-LFSR. While the LFSRs generate their random values, the cell translates any matches between the generated random numbers and received the Cartesian locations into flags stored into 4-bit registers, which are later decoded by X-DMUX and Y-DMUX. Here, a match means there is a distal connection established between the two cells. It is important to mention here that following such an approach makes the process of forming distal connection probabilistic, while in HTM network it is deterministic. However, in HTM, the cells that are currently active form connections with a subset (typically 50\%) of the cells that were active in the previous time step, and in our design this is achieved naturally through our adopted probabilistic approach. Now, in order to estimate the likelihood of matching between distal segment addresses (randomly generated) and the addresses of the active cells,~\eq{match} can be used, where $n_{sd}$ is the maximum number of synapses in a distal segment. Let the distal segment size for a given cell be 256. Given 961 mini-columns with 40 actives at each time step, there is a 0.847 likelihood that at least 20\% of the generated random address matches those of the previous active cells. This likelihood can be significantly increased beyond 0.95 when the segment size\footnote{Increasing the distal segment size cost more cycles to generate more random addresses and additional memristor devices for each new added synapse.} is increased, as shown in~\fig{match_prob}. 
\begin{equation}
\centering
P_{match} = frac{\sum\limits^{n{_w}}_{i=10} {n_w \choose i} \times {n_c - n_w \choose n_{sd} - i}}{{n_c \choose n_{sd}}}
\label{match}
\end{equation}

After finishing the matching process and activating the X-DMUX and Y-DMUX, all the possible combinations of 16 X-addresses and 16 Y-addresses are achieved through the AND gate array. The output of logic '1' for an AND gate, let's say gate number 5, may indicate an active cell in the location ($x$=1 and $y$=5). The output of the AND gate enables the corresponding 'green' 2-bit register to load the Z-address, and this represents the cell distal synapse that is currently connected to an active cell at time $t-1$, whereas the previously formed distal synapses are stored in the 'blue' 2-bit register. However, once the registers are loaded, they are compared and the results are relayed to the distal segment memristors (only when evaluating the cellular activities detected by distal segment). For the distal segment unit, this cell architecture leverages the union propriety of the SDR representation to considerably reduce the cell architecture complexity. The main concept behind the union property is storing several patterns using one representation. This can be translated into having one universal distal segment for each cell rather than multiple of them. The universal segment grows as the cell learns more temporal information. It is important to mention here that merging the segments can increase the possibility of false triggering of cell segments and incorrect predictions. However, this is less likely to happen if we limit the number of patterns ($M$) a segment can learn, while setting the number of mini-columns and cells to be large enough. For instance, in this work, we used 961 mini-columns with 4 cells each. If we stored 30 patterns in a segment and set the matching threshold for any two given patterns to 5, according to~\cite{ahmad2016neurons}, the probability of a false match is $6.408\times10^{-14}$ as calculated using~\eq{match_eq}.
\begin{equation}
\label{match_eq}
    P_{fp} =\left[1 - (1- \frac{n_w}{n_c})^M \right]^{n_w}
\end{equation}
\begin{figure}[t]
\begin{center}
\includegraphics[width=0.48\textwidth]{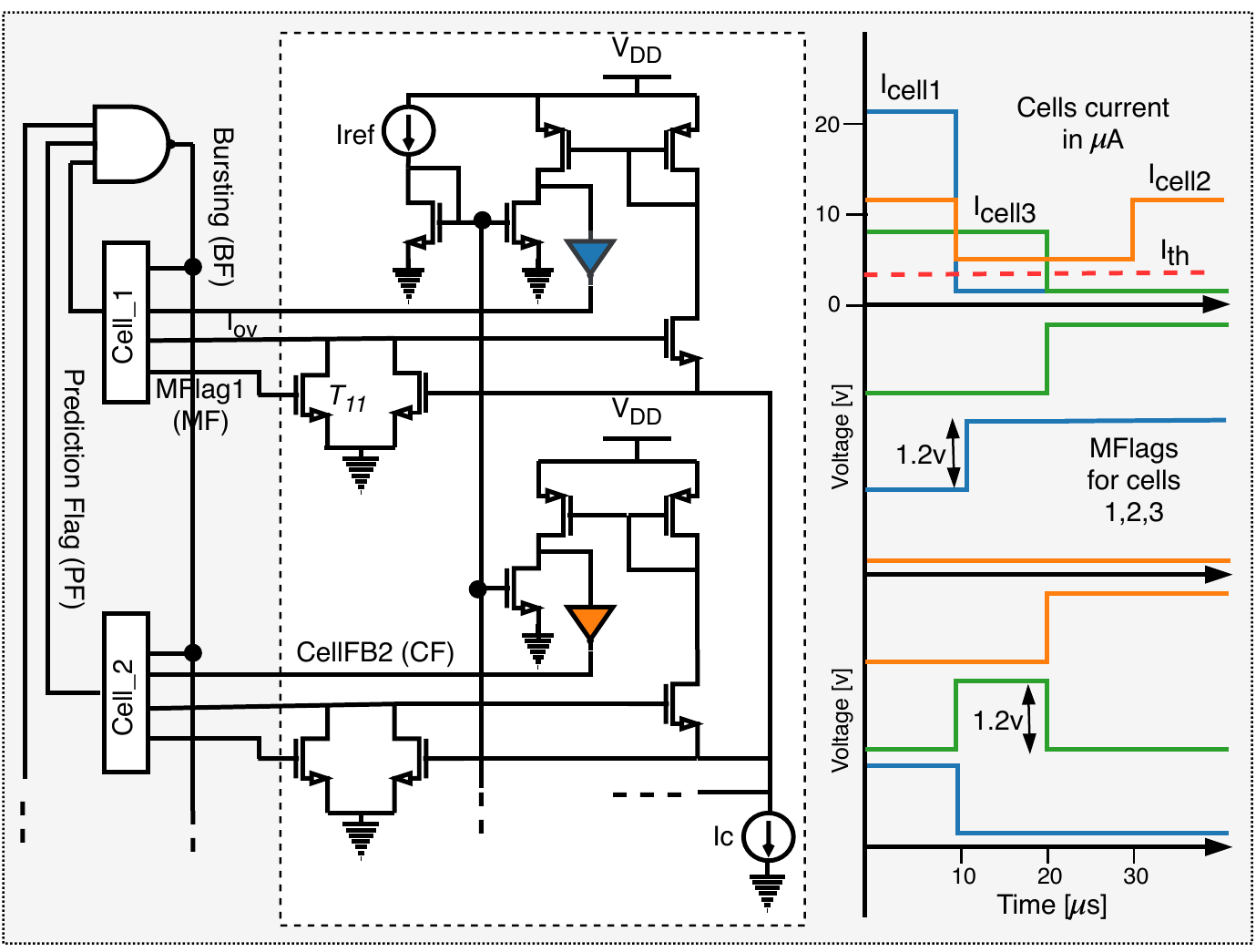}
\caption{(left) Competitive circuit that enables the cells within one mini-column to interact with each other when a massive firing activity takes place in the mini-column. (right) Waveform diagram illustrating the competitive circuit for the cell to select the best matching cell.\vspace{0mm}}
\label{all_cells}
\end{center}
\end{figure}

The output current that is collected at the distal segment bitline is received by the current comparator unit. Then, the current gets mirrored\footnote{The bursting mini-columns' cells generate additional current that should add up to the segment current during the evaluation step. However, the bursting mini-columns are evaluated globally, at the region level, and their contribution to the segment activation is done through $I_{burst}$.} to be compared with two reference currents: active threshold and learning threshold. If the segment current is more than the active threshold, the segment is set to be in an active state and consequently the cell state changes to predictive for the next time step. On the contrary, current less than the active threshold and more than the learning threshold, marks the cell as a matching cell. A matching cell has a high probability to be selected to represent the input when bursting takes place. It is important to mention here that the prior discussed operations are carried out within the cells, but running the temporal memory successfully also requires the cells within the mini-columns to interact with each other to identify whether bursting is necessary. If bursting takes place in a mini-column, all the cells within the mini-column are set to be active and one cell is selected to learn the current input pattern. Typically, this is done either by selecting the best matching cell or least used cell. The former occurs only when a cell has a sufficient number of potential synapses that are connected to active cells in the previous time step i.e. a cell has a matching segment. Choosing the best matching segment involves selecting the cell with the highest matching level (distal current). This implies mirroring all cells' output current to another unit, namely the competitive circuit (a modified current based winner-take-all circuit originally proposed in~\cite{lazzaro1989winner}), so that the cell with the highest output current is chosen (see~\fig{all_cells}-(left)). In the case when there are no matching segments, the least used cell is chosen as a winning cell. Selecting the least used cells is done via selecting the cells with the least number of distal segments. Since this implementation deals with one universal merged segment, a counter in the cell is used to monitor the flags of added segments and consequently the number of merged segments in each universal one. \fig{all_cells}-(right) demonstrates the operation of the cells competitive circuit. Here, three cells are competing to select the best matching cells. Two scenarios are considered. The first of which (interval 0-10 $\mu$s), all cells have high overlapping current (all MFlags = '0') so that they are in competition. Since cell$_1$ has the highest overlapping current, it is selected as a winner. In the second scenario (interval 10-40 $\mu$s), cell$_1$ has less current than the `Active Threshold', for this reason it is excluded from the competition. This is accomplished via switching $T_{11}$ to an ON state, and this eventually blocks cell$_1$ current which is mirrored to the WTA circuit.

%$$$$$$$$$$$$$$$$$$$$$$$$$$$$$$$$$$$$$$$$$$$$$$$$$$$$$$$$$$$$$$$$$$$$$$$
\section{Synthetic Synapses Representation} 
The cells in the HTM network interact with each other during the temporal memory phase. This interaction is essential to enable the network to predict the upcoming events. As alluded to earlier, the cells' interaction is enabled through the distal segments which are established and evolved while learning temporal information. In hardware, this translates into thousands of interconnects that are continuously changing in their conductivity level and locations. Due to the fact that interconnects in VLSI systems are rigid in nature and do not support this level of reconfigurability, memory units can be used to virtually formulate these connections and to describe their strength as in~\cite{Goldberg2001, zyarah2015reconfigurable}. Although such an approach is effective as it endows the network with the necessary dynamic to learn spatial and temporal information, it does not suit edge devices which have stringent area and energy constraints. Thus, we are presenting the SSR communication scheme that heavily relies on random generators and memristor devices rather than conventional memory units to form synaptic connections and to define their growth levels. This results in significant savings in terms of resources and energy consumption.

\begin{figure}[t]
\begin{center}
\includegraphics[width = 0.48 \textwidth]{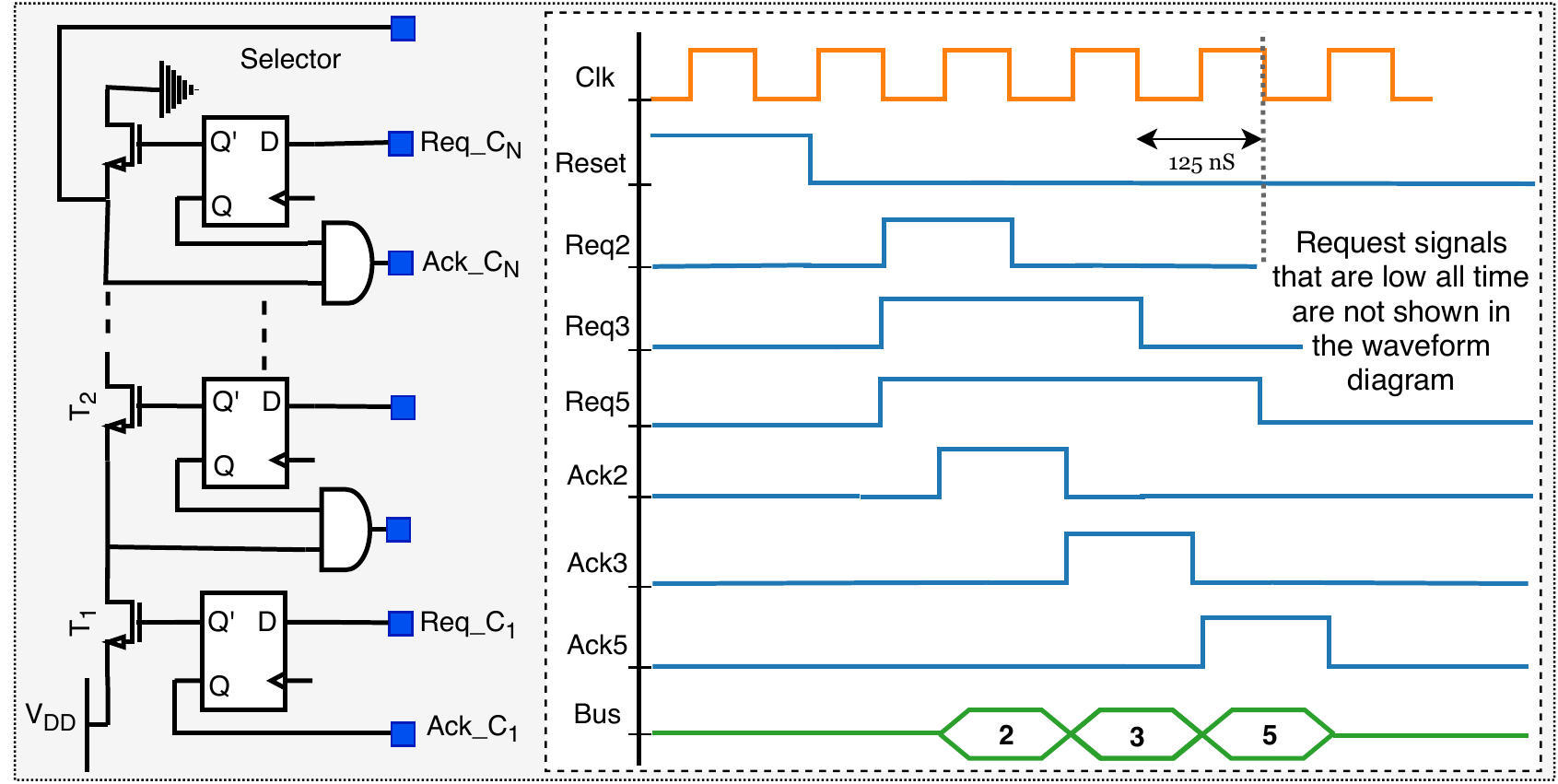}
\caption{(left) A synthetic synaptic arbiter (SSR) circuit consisting of buffers to store the simultaneous requests from the winning mini-columns, a series of nMOS pass transistors to monitor the status of the individual mini-column's requests, and a feedback circuit to clear mini-column requests once served. (right) Waveform diagram demonstrating a part of the SSR operation while processing several concurrent requests sent from several mini-columns located within the same row of the HTM region.\vspace{0mm}}
\label{arbiter}
\end{center}
\end{figure}

Two aspects associated with the SSR are addressed in this work: forming synaptic connections using LFSRs (discussed earlier in section \ref{htm_cell_section}) and controlling the data transfer among cells through regulating the access to the H-Tree bus. Considering the same HTM system with $A^{t-1}$ active cells in the previous time step and $A^t$ active cells in the current time step, during the temporal memory phase, every cell in the network with enough strong connection to $A^{t-1}$ cells can be depolarized for the next time step and become predictive. The challenge here is how to transfer the $A^{t-1}$ cells' addresses to all other cells in the network efficiently. Let all the mini-columns with active cells at time $t-1$ place a request at the input of the outgoing tri-state gates (see~\fig{htm_soc}). Then, each set of tri-states belonging to the same row are activated simultaneously through the selector. When a row is selected all its tri-state buffers associated with the mini-columns are activated, allowing the mini-columns to send requests to the arbiter and to receive acknowledgements. The arbiter circuit is shown in~\fig{arbiter}-(left). It comprises of buffers, a series of nMOS pass transistors, and a feedback circuit. The buffers are used to store the simultaneous requests from the selected mini-columns. The series of pass transistors are used to monitor the status of the individual mini-column requests, whereas the feedback circuit is used to acknowledge the mini-columns after their requests are served. In~\fig{arbiter}-(right), a waveform diagram illustrates the operation of the arbiter, selector, and other units in the developed system while processing information sent from a row with 5 mini-columns. Initially, all the winning mini-columns' (in this example: 2, 3, and 5) requests are directed toward the arbiter and stored in the buffers (DFF). When the DFF-3, for instance, receives $Req_3$, it waits in a queue until $Req_2$ is served. Once $Req_2$ is served, the voltage drop at $T_2$ drain will be high. This will trigger the feedback circuit to send $ack_{3}$ signal to mini-column 3, which in turn clears its request and broadcasts the address of its active cell(s). Serving the requests of all the active cells in the HTM region leads to a latency given by: 
\begin{equation}\label{cell_to_cell}
t_{cc} = \sum\limits_{i=1}^{\sqrt{n_c}}(\sum\limits_{j=1}^{\sqrt{n_c}} {\Lambda}[i][j] + 1)
\end{equation}

Recall that the SSR conveys the same concepts of the AER and the enhanced AER, but it is designed to serve intra-chip communication while offering the following advantages: 
\begin{itemize}
    \item In AER, the neuron potential duration must be $\approx$500 times more than the event duration for transmission to time-multiplex the transmission channel~\cite{mahowald1992vlsi}. There is no need for such a constraint in the SSR. 
    \item The enhanced AER demands memory units on both sides, sender and receiver, to hold neuron addresses that are virtually connected (connecting 32x32 cells requires 20Mb RAM~\cite{Goldberg2001}). For a sparse network like the HTM, this is very overwhelming in terms of memory usage. However, in the SSR, the addresses are generated rather than stored. This serves two advantages: smaller storage units are used and random selection is achieved. 
    
    %Memory usage per cell = 10+(1024-1)*10+ (1024-1)*10 + (1024-1)*16 = 36838 bits
    % To store only addresses (no permanence), memory usage = 36838 - (1024-1)*16 = 20470 bits 
    
    \item The SSR is synchronous and its capacity, the maximum rate of sample transmission (considering the worst case scenario and the adopted network architecture), is 4MSamples/sec. In the AER case, its capacity for SNN with approximately the same network size is 2.5MSamples/sec ~\cite{boahen1998communicating}.

    %\item Signal integrity and collision 
    
    % ... The other possible approach is to use encoders at the sender side and a decoder at the receiver side. Such approach turns to be effective if the number of neurons are limited. Because it does not make sense to build an encoder for 1024 input unless some sort of hierarchical encoding and decoding is used! Or in dimension, the encoding is going to be for 32 input for x-axis and 32 for y-axis. The downside of encoding and decoding approach is that in HTM, the destination address is not one location shared between several neurons just as in fully connected network. Rather, the destination address may be decoded to multiple location at the receiver side! The question he is how costly it is not build x-y decoder in each cell? is it more costly than using 40 memory locations?
    
    % \item The AER takes the advantage of the disparity in event and action potential duration to share the data buses and save resources..

    % \item In two dimensional space, we need to encode one event at a time! This will slow down the process. Here, I'm suggesting to included the y-dimension implicitly though packet orders. In this manner, we have to go over the array row by row! and I will need a row enable pointer or just a shift register next to the rows!
    
    \item The SSR uses priority arbiter, which applies a queuing mechanism to access the H-Tree (or channel) bus, whereas AER utilizes an arbitration mechanism to access the channel. The latter is known to lengthen the communication cycle period and reduce channel capacity~\cite{boahen1998communicating}. 
    
    % \item In AER, we are transferring the current neuron events at time t, whereas in HTM case, we need to transfer the events that occurred at time t-1

    \item The AER is deemed an effective approach for inter-chip communication, where neuronal information is communicated by means of encoded events. At the targeted destination, the encoded events are typically decoded and routed to the proper accessible neurons. The encoder size here is highly dependent on the number of neurons, whereas in the SSR, the decoding process complexity is defined by the number of synapses associated with the targeted neurons. This property is extremely beneficial for sparse networks like HTM.
    
    \item The enhanced AER offers better flexibility in updating the synaptic connections individually. The opposite is true for the SSR, in which changing the seeds of LFSRs enables the cell to form a new set of synaptic connections.
\end{itemize}

% ==========================================
\section{Experimental Methodology}
\begin{figure}[t]
\begin{center}
\includegraphics[width = 0.35 \textwidth]{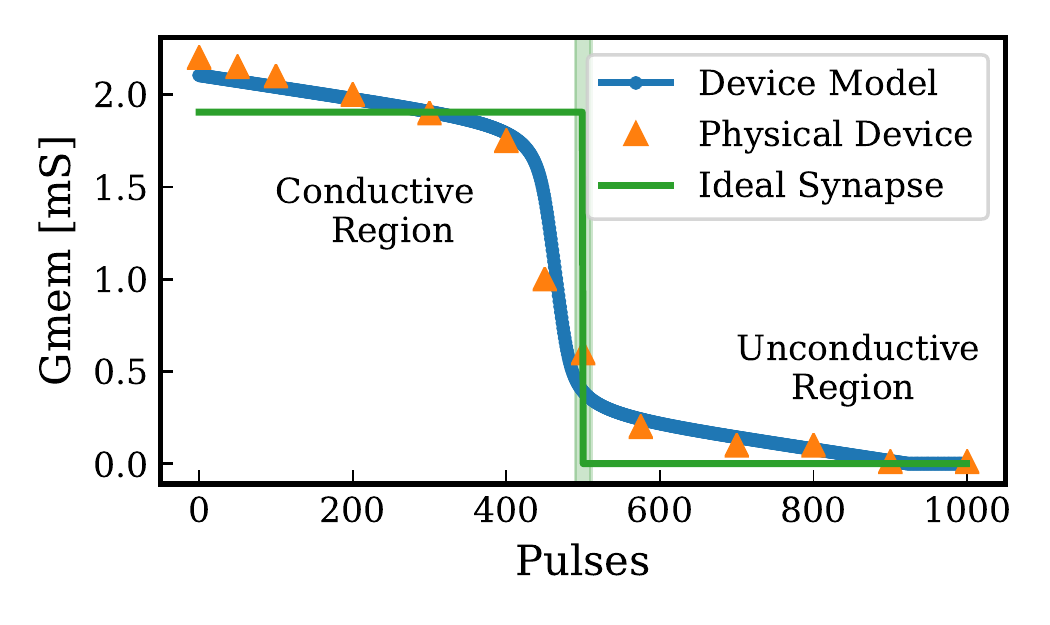}
\caption{Fitting the memristor model to the physical device behavior while modulating the device conductance with a train of pulses.}
\label{mem_mapping}
\end{center}
\end{figure}

In order to assess the performance of the proposed mixed-signal HTM system, two models are created. The first is a golden model (HTM-SW) that runs the HTM system without any constraints. This model is used to find the optimal network performance for a given task. The second model (HTM-HW) is an emulation of the hardware design under predefined circuit constraints. Here, the circuit constraints\footnote{Memristor device non-idealities considered during the simulation are: 10\% cycle-to-cycle variability (memristor resistance) and device-to-device variability (write variation).} are achieved after the individual components of the design are simulated and verified within Cadence Virtuoso environment. Prior to that, all the digital units are verified for functionality in Cadence SimVision. During simulation, the supply voltage is set to 1.2v and the system clock is at 8MHz. The system also has 128MHz high-speed clock to drive the LFSRs of the cells. When it comes to emulating HTM synaptic connections' strength, a representative non-linear Verilog-A memristor model~\cite{kvatinsky2015vteam} with a modified Z-window function~\cite{zyarah2019neuromemrisitive} is utilized\footnote{$\tau$, $\delta$, $k$, and $p$ are constants to control the window function shape. The nominal values used in this work are: $\tau$=200, $\delta$=0.5, $k$ =1, and $p$=4}. The device conductance changes as a function of the state variable, $w$, is described in~\eq{mem_eq} and~\eq{mem_eq2}\footnote{$k_off$, $k_{on}$, $\alpha_{on}$, and $\alpha_{off}$ are constant, and $v_{off}$ and $v_{on}$ are the memristor threshold voltage.}, where $D$ is the device thickness, and $G_{on}$ and $G_{off}$ define the memristor conductance limits. Emulating the synaptic behavior of HTM using memristors turns out to be challenging. This is because the synapses in HTM are binary in nature, i.e. they exhibit the same properties if they are above the permanence threshold regardless of the synapse's growth level and vice versa. In 2017, Jiang et al. proposed a memristor device to implement the $k$-nearest neighbour algorithm and that exhibits properties required for HTM~\cite{jiang2017rram}. \fig{mem_mapping} illustrates the experimental behavior of the physical device as a function of the applied pulses, fitted to the memristor model. Here, it can be observed that the memristor has minor changes in conductance level on either side of the permanence threshold (highlighted in green), while the changes are extreme in the middle. To some extent, this captures the binary nature of the ideal synapse in HTM. It is important to mention here that in order to optimize the HTM system performance and maintain low power consumption, the following assumptions were made: 1) the memristor device exhibits semi-symmetrical behavior when switching from low/high conductance to high/low; 2) the memristor  device offers fast switching speed and high conductance range.~\tb{mem_para_table} shows all the device parameters used for proximal and distal synaptic connections.

\begin{table}[h!]
\caption{The memristor device parameters used in the mini-column and cell designs.\vspace{-2mm}}
\label{mem_para_table}
\setlength\tabcolsep{3 pt}
\renewcommand{\arraystretch}{1.3}
\begin{center}
\begin{tabular}{|l|cc|}
  \hline                       
\rowcolor{Gray}
\textbf{Parameter} & \textbf{Value [mini-column]} & \textbf{Value [cell]} \\ \hline
Proximal memristor range & 150k$\Omega$ - 10M$\Omega$  & 150k$\Omega$ - 10M$\Omega$\\ 
Memristor threshold      & $\pm$0.95v & $\pm$0.95v \\ 
Full switching pulses    & 51 & 51 \\ 
Training voltage        & 1.1 v & 1.1 v\\ 
Sense memristor range   & 20k$\Omega$-80k$\Omega$ & -\\ \hline
\end{tabular}
\end{center}
\end{table}

\begin{equation}
G_{mem} = \frac{w}{D} \times G_{on} + (1 - \frac{w}{D}) \times G_{off}
\label{mem_eq}
\end{equation}
\begin{equation}
\frac{\Delta w}{\Delta t} = 
\begin{cases}
k_{off}.\Big(\frac{v(t)}{v_{off}} - 1\Big)^{\alpha_{off}}.f_{z}(w),&0 < v_{off} < v \\
0, &v_{on} < v< v_{off} \\
k_{on}.\Big(\frac{v(t)}{v_{on}} - 1\Big)^{\alpha_{on}}.f_{z}(w),&v <v_{on} < 0
\end{cases}
\label{mem_eq2}
\end{equation}
\begin{equation}
\label{mem_eq3}
    f_z(w) = \frac{k[1-2 (\frac{w}{D} - \delta)]^p}{e^{\tau (\frac{w}{D} - \delta)^p}}
\end{equation}

%###################################################
\section{Results and Discussion}

\subsection{Time-Series Prediction}
The prediction accuracy of the proposed HTM system is evaluated using real-world streaming data. Given an input dataset of length $n_n$, where each data point presented to the HTM system at time $t$ is represented by $y^t$, while the corresponding predicted value is given by $\hat{y}^t$, the mean absolute percentage error (MAPE) can be computed as in~\eq{mape}.

\begin{equation}
\centering
MAPE = \frac{\sum\limits^{n_n}_{t=1} |y^t - \hat{y}^t|}{\sum\limits^{n_n}_{t=1} |y^t|}
\label{mape}
\end{equation}

\begin{figure}[h!tb]
\begin{center}
\includegraphics[width = 0.35 \textwidth]{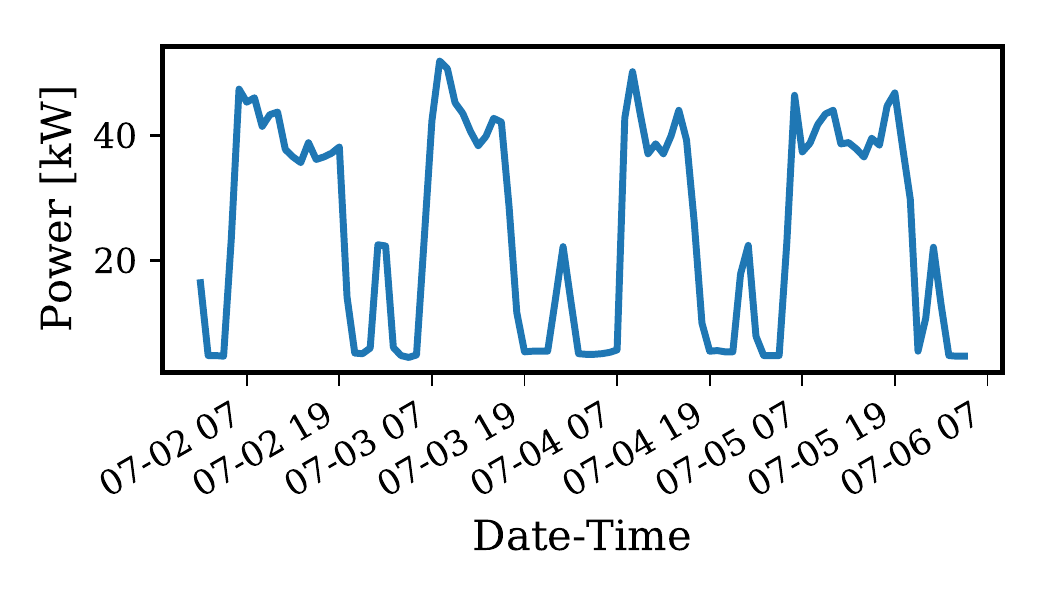}
\caption{A snapshot of the power consumption of the Hot-Gym dataset~\cite{hotgym} recorded every hour over approximately 4 days.}
\label{hot_gym}
\end{center}
\end{figure}

\begin{figure*}[h!tb]
\centering
\subfigure{\includegraphics[width=60mm, height=44mm]{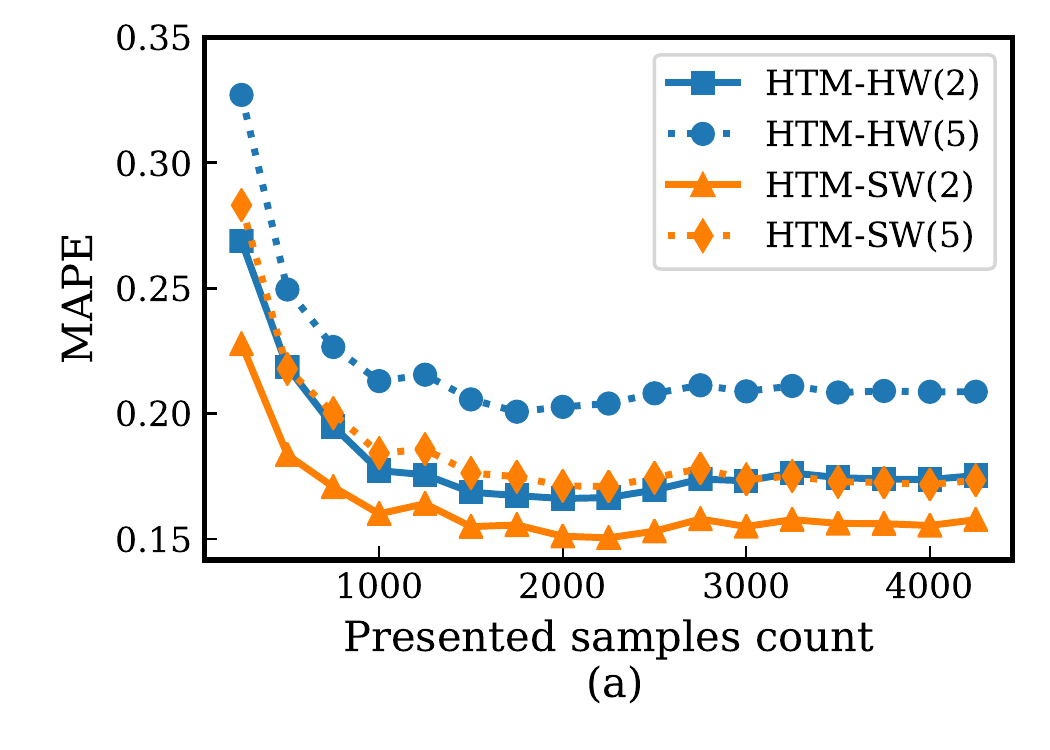}}
%\hspace*{-5em}
\subfigure{\includegraphics[width=60mm, height=44mm]{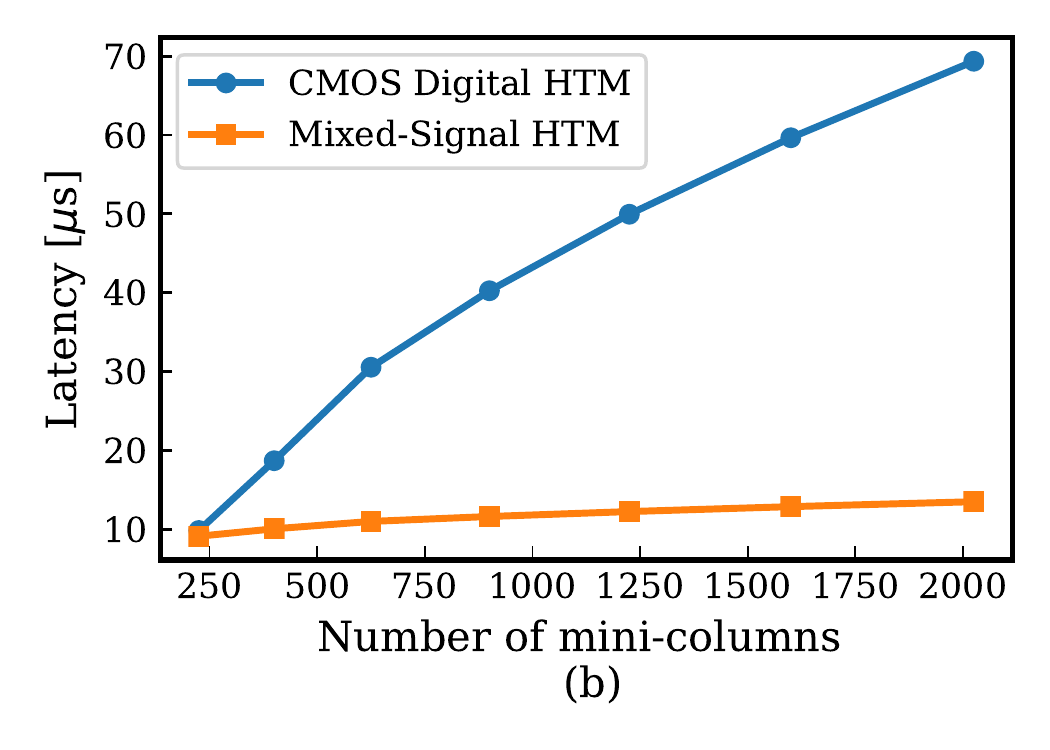}}
%\hspace*{-5em}
\subfigure{\includegraphics[width=60mm, height=44mm]{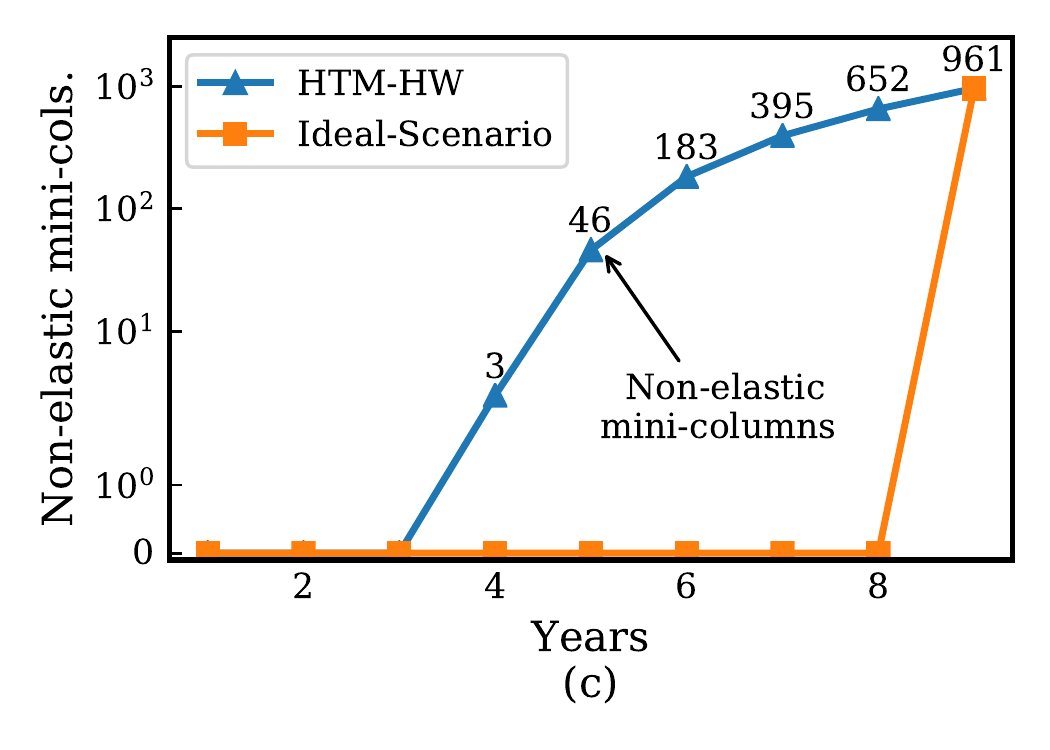}}
\vspace{-7mm}
\caption{(a) MAPE for predicting the power consumption in a gym for the next 2 and 5 hours using HTM software (HTM-SW) and HTM hardware (HTM-HW) models. (b) Latency of the digital and mixed-signal HTM as a function of the network size, given by the number of mini-columns. (c) Elasticity (lifespan) of the overall HTM mini-columns in the ideal and real-world scenarios. \vspace{-2mm}}
\label{metric}
\end{figure*}

\fig{hot_gym} illustrates a snapshot of the Hot-Gym dataset~\cite{hotgym}, the power consumption in a gym, over a small period. The power consumption is recorded at every hour for 4 months (total samples count = 4390). Here, the HTM system is used to predict the power consumption for the next 2 and 5 hours. Initially, the golden software model, HTM-SW, is used in the prediction. Then, the same prediction is made using the HTM-HW model\footnote{HTM-HW model is also benchmarked using other datasets such as NYC-Taxi~\cite{trip_record}. The achieved MAPE for the 2nd and 5th order predictions are 0.0996$\pm$0.0014 and 0.156 $\pm$ 0.0084, respectively.}.~\fig{metric}-(a) shows the accumulated MAPE recorded at every 250 samples. It can be seen that the initial value of the MAPE is really high, but over time it decreases as the network learns patterns and uses the acquired knowledge to make valid predictions in the future. However, the overall MAPE of the software model, assuming the first 500 samples presented to the network are dedicated to learning, is calculated to be 0.154 $\pm$ 0.0014 (0.171 $\pm$ 0.002 for 5-step prediction), while the hardware equivalent is 0.174 $\pm$ 0.002 (0.205$ \pm$ 0.0046 for 5-step prediction). This degradation may be attributed to the unsymmetrical characteristics of the memristor devices leading to disparity in the network learning and forgetting rate. Applying the union property to the distal segment and forming its distal synapses using LFSRs might have negative consequences as well, especially when making higher order predictions.

%-----------------------------------
%> There is extra faces in cmos
%> 
\subsection{Latency}
The latency is measured as the time required for the HTM network to process an SDR input generated by the encoder. In this context, HTM processes SDR inputs of the Hot-Gym dataset, where each input is encoded with 512-bit binary vector. The spatial pooler and temporal memory phases here are performed simultaneously\footnote{Spatial pooler and temporal memory operate simultaneously when the H-Tree bus is exploited by either of them.} and in a pipelined fashion to minimize the latency, which is estimated to be 11.64 $\mu$s.~\fig{metric}-(b) shows the latency of the CMOS digital HTM (system clk = 100MHz) and the proposed mixed-signal HTM (system clk = 8MHz) as a function of the network size, given by the number of mini-columns. One can notice that the latency in the digital HTM is always higher than the mixed-signal counterpart. This can be attributed to several reasons. The first is the need for the initialization phase in the digital HTM design to set the synaptic connections' permanences, particularly the proximal synapses, prior to receiving any input. The initialization of the synaptic connections' permanence is achieved for free in the mixed-signal design as the memristors after the formation process have random conductance with Gaussian distribution~\cite{prezioso2015training}. Second, tuning the synaptic connections, proximal or distal, is performed simultaneously at the cell and mini-column levels, but within them it is sequential because the permanence values are stored in distributed SRAMs, where the read/write operations take several clock cycles. In the mixed-signal design, on the contrary, the tuning process is performed concurrently even within the mini-columns or cells and usually it takes two clock cycles. Finally, in the digital HTM, the winning mini-columns that represent the input are decided in a sequential fashion to cut down the resource cost and power consumption. This in turn translates to longer latency that is proportional to the number of mini-columns. In the mixed-signal design, a WTA circuit~\cite{zyarah2019neuromemrisitive} is used, which processes all the inputs concurrently.

%------------------------------------
\subsection{Network Reliability and Lifespan}
The memristor device write endurance, which is the number of times a memory cell can be overwritten successfully, turns out to be a crucial factor in determining network sustainability for learning. The memristor devices, particularly oxide-based devices, have a typical endurance range between $10^6 - 10^{12}$~\cite{coll2019towards}. This low endurance reduces the network reliability for online learning and continuous adaptation especially when the network is densely connected and all neurons need to be updated continuously. For the HTM network, this is not the case as cell/mini-column activities are sparse in nature and the learning is confined only to the active ones. This feature endows the network longer elasticity (lifespan) in comparison to other networks. In order to estimate the elasticity of mini-columns in the HTM network, we need to estimate their successful training rounds ($L_r$) and likelihood of activation, as given by~\eq{life_period}, where $E_d$ is the memristor device endurance. 

\begin{equation}\label{life_period}
L_r = \frac{E_d \times n_c}{n_w}
\end{equation} 

\begin{figure*}[h!tb]
\centering
\subfigure{\includegraphics[width=60mm, height=44mm]{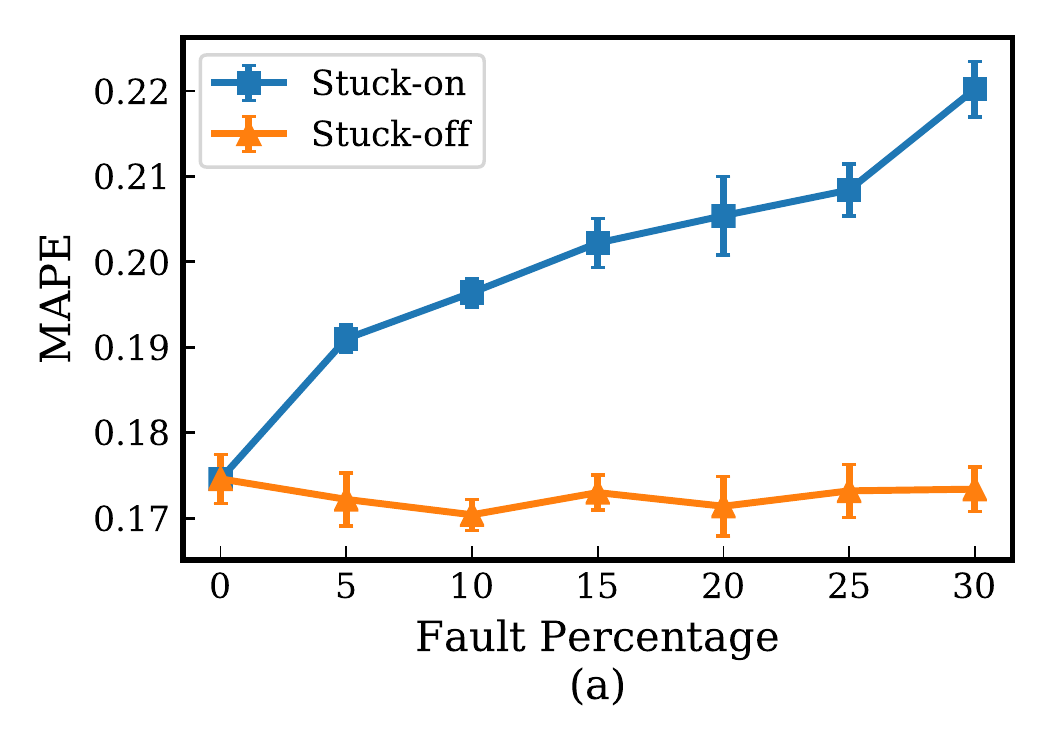}}
%\hspace*{-5em}
\subfigure{\includegraphics[width=60mm, height=44mm]{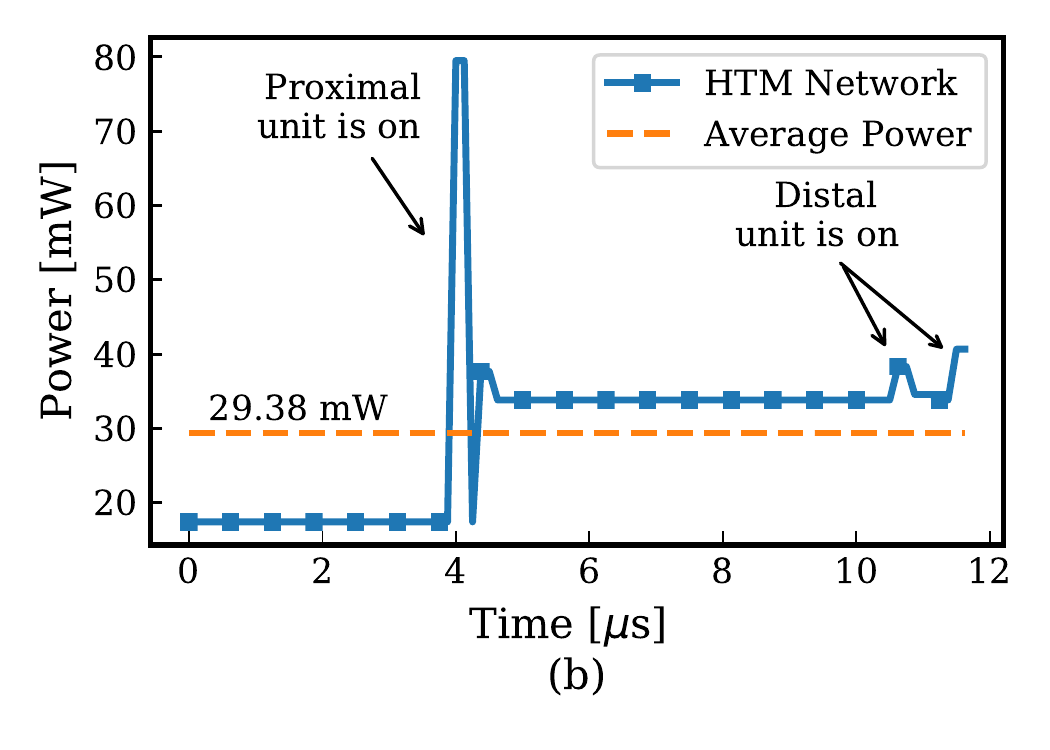}}
%\hspace*{-5em}
\subfigure{\includegraphics[width=60mm, height=44mm]{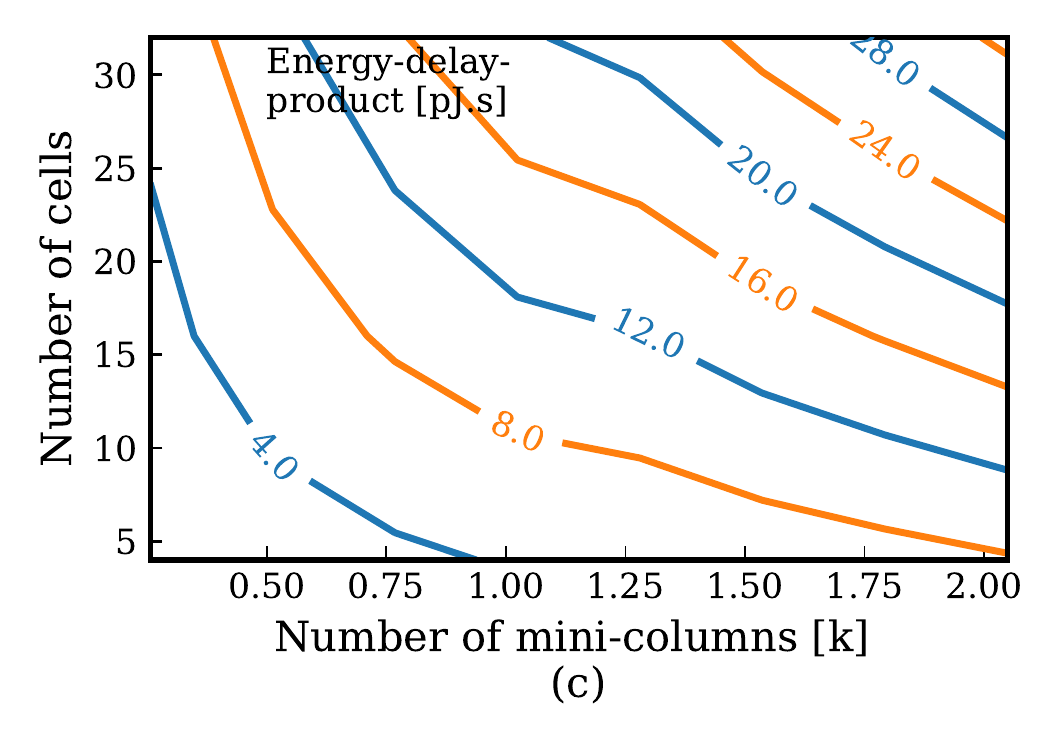}}
\vspace{-7mm}
\caption{(a) The MAPE of the HTM-HW predicting two steps ahead in time for the Hot-Gym dataset while experiencing various types of stuck-at faults. (b) The total power consumption of the developed HTM system as it processes and predicts time-series data from Hot-Gym dataset. (c) Contour of energy-delay-product for the developed HTM system as a function of network size.\vspace{-2mm}}
\label{metric2x}
\end{figure*}

In the ideal scenario, mini-columns in the HTM network are activated with equal likelihood by patterns detected at the proximal segments. Thus, the number of successful learning rounds that can be made, given $E_d$ = $10^9$, $n_c$=961, and $n_w$=40, is $240 \times 10^8$. This is equivalent to $\approx$8 years of successful continuous learning performed at a rate of 10ms. Comparatively, this is $\approx$24 times more than a conventional network with no sparse activities, and X\footnote{X is not specified here because it is highly affected by the input and the encoding approach.} times more than the SNNs\footnote{SNNs are usually trained with spike-time-dependent-plasticity (STDP) rules. STDP requires neurons to be tuned based on the time difference between the pre and post synaptic neurons' spikes.}. In spite of the fact that SNNs are asynchronous and sparse in nature, usually their neurons fire and their synaptic connections are tuned multiple times while processing a single input. This is because each input is stochastically encoded as a stream of spikes. However, the previous comparison hypothesizes that the HTM mini-columns' activations are perfectly regularized by incorporating the homeostasis plasticity mechanism (or boosting). In real-world scenarios, this is not the case, because the mini-columns' activations are highly affected by input space statistics.~\fig{metric}-(c) is an example demonstrating an estimation of the developed system elasticity (lifespan) for the Hot-Gym dataset. Here, we see that after year 4, a gradual loss in mini-columns' elasticity starts to occur. Even after 8 years of work, $\approx$309 mini-columns are still elastic and have the capability to acquire new information. However, the overall network performance at that time would be limited. 

%--------------------------------------
\subsection{Device Failure and Network Robustness}
There are various types of memristor defects that may affect network performance, and usually they occur due to process variation~\cite{kannan2014detection, romero2019training}. Examples of device defects are ageing faults, endurance degradation faults, switching delay faults, and stuck-at faults~\cite{kumar2017oxygen, ravi2018fault}. Here, we will emphasize on the stuck-at fault as it is ubiquitous and has high impact on network performance~\cite{ravi2018fault}. Two types of stuck-at faults are studied. The first of which investigates the impact of stuck-on (high-conductance state) on HTM system performance while making two step ahead prediction for the Hot-Gym dataset. The second focuses on the stuck-off (low-conductance state) effect.~\fig{metric2x}-(a) illustrates the averaged MAPE over 5 runs for the HTM-HW prediction as a function of the faulty\footnote{The fault is applied to the mini-columns' proximal connections and SDR classifier weights.} device percentage for the aforementioned cases. It can be seen that the stuck-off fault has a positive marginal impact on the network performance as it leads to an increase in the network sparsity level. In contrast, the stuck-on increases the MAPE by 1.7\% and it can go up to 4.9\% when the fault percentage is 30\%. This degradation in performance arises from the fact that the SDR classifier is implemented using a softmax classifier with weighted synapses realized using a memristive crossbar. Having 10\% of stuck-on fault in crossbar means on average, every row and column in the crossbar has 55 and 344 defected devices, respectively. This eventually makes the softmax classifier output nodes unable to distinguish various pattern activities and fire excessively. 

During the fault analysis, it is also found that applying the fault solely to the spatial pooler results in a marginal change in the system performance. This is because each input sample presented to the HTM is spatially represented by a small population of active mini-columns, and having a slight change in the representation pattern, which may result from the fault, has very low impact. Furthermore, using $k$-winner mechanism mitigates the changes that may occur in spatial patterns. 

\begin{table*}[h!tb]
\caption{A comparison of the proposed HTM system with previous work. One may note that these implementations are on different substrates, \textit{thereby this table offers a high-level reference template for HTM hardware rather than an absolute comparison \vspace{-2mm}}.}
\label{HardwareAnalysis}
\setlength\tabcolsep{3 pt}
\begin{center}
\begin{threeparttable}
\begin{tabular}{|c|ccccc|}
\hline                      
\rowcolor{Gray} 
\textbf{Algorithm} & \textbf{Memristive HTM \cite{krestinskaya2017hierarchical}} & \textbf{PIM HTM \cite{zyarah2019neuromorphic}} & \textbf{Digital HTM~\cite{li2016hardware}} & \textbf{PE HTM}~\cite{li2019design} & \textbf{This work} \\ \hline 
  Task   & Classification    & Classification\&Prediction & Prediction   & Image recognition & Classification\&Prediction  \\ 
  Operating Frequency  & - & 100MHz &  100MHz  & 100MHz & Dual 8-128 MHZ\\ 
  Proximal Segment Size  & 9 & 16 & 1 & 40 & 31\\ 
  Distal Segments x Size    & - & 5x10 & - & 12x16 & Shared 256\\ 
  Total Power consumed & 13.34mW\tnotex{tnote:robots-a2}  & 417mW & 516mW\tnotex{tnote:robots-a1} & 4.1W & 29.38mW\\ 
  Dataset &	AR, TIMIT, \& ORL &	MNIST & MNIST & KTH & MNIST\tnotex{tnote:robots-a4} ~\& Hot-Gym\\
  Mini-columns x cells & 25xX\tnotex{tnote:robots-a3} & 	100x3 & 400x2 & 2048x32 & 961x4\\ 
  Latency (ms) &  -      &     0.0057         &  0.0045  & 6.04 & 0.0116   \\ 
  Technology node     &  TSMC 180nm      &     TSMC 65nm         &  Nangate 45nm       & GF 65nm  & IBM 65nm   \\ \hline
  \end{tabular}
      \begin{tablenotes}
      \item\label{tnote:robots-a1} In~\cite{li2016hardware}, the power consumption is reported for a single processing element (PE) without considering the register files. Thus, we linearly scaled the power for an HTM network of size 400x2.
      \item\label{tnote:robots-a2} In this reference, the temporal memory power is reported for single pixel processing. This value is multiplied by the total number of mini-columns to estimate a total power of an HTM region with 25 mini-columns with one cell each.
      \item\label{tnote:robots-a3} X denotes unknown number of cells.
      \item\label{tnote:robots-a4} Further details about MNIST results are provided in~\cite{zyarah2019neuromemrisitive}.
    \end{tablenotes}
\end{threeparttable}
\end{center}
\vspace{-5mm}
\end{table*}

\subsection{Power Consumption and Distribution}
The average total power consumption of the developed HTM system while predicting time-series data from the Hot-Gym dataset is estimated to be 28.94mW and 29.38mW\footnote{The approach used to estimate the power consumption is described in our previous work~\cite{zyarah2019neuromemrisitive}.} when the online learning is enabled. The high power consumption during the training is due to the use of high voltage (memristor training voltage $\approx $1.1v) and extra clock cycles to modulate the memristor devices.~\fig{metric2x}-(b) demonstrates the estimated total power consumption over time. Initially, 17.18mW is consumed while transferring the input SDRs through the H-Tree\footnote{The H-Tree structure might be buffered with full-swing and reduced swing buffers, proposed in~\cite{asgari2004low}, to minimize the power consumption further.} to the mini-columns and establishing the proximal synaptic connections, which take place in simultaneous fashion. The power then abruptly increases due to the activation of the proximal segments to compute the mini-column overlap scores. Once the winning mini-columns are selected, the spatial pooler learning phase starts, in which the memristors associated with proximal synapses are modulated. Meanwhile, the prior active cells' addresses are routed to each cell in the winning mini-columns to compute their distal segments' overlap scores. Computing distal segments' overlap scores give rise to another abrupt increase in the power consumption (at time $\approx10.6 \mu$s). However, this increase is much smaller than the one occurred while computing the overlap scores of the mini-columns. This is because computing the cells' overlap scores is confined only to the cells within the winning mini-column while other cells are disabled through clock-gating. After computing the cells' overlap scores, the cells of the winning mini-columns locally compete to represent the input contextually. The selected active cells form the lateral connections with the neighboring cells and tune their distal connections accordingly. One may observe from the previous discussion that tuning and computing the overlap scores here turn to be the most power-hungry operations as there are more than 45.15k
% Synapses counts = 60 x 256 + 961 x 31 = 
synapses involved in the network computations. One possible way to minimize the power is to modify network size or segregate the above operations into multiple stages at the mini-column or cell levels, but this will be at the expense of increasing the overall network latency.~\fig{metric2x}-(c) illustrates the contour of energy-delay-product measured in pJ.s which can be used to pick the optimal network architecture for a given power consumption and latency requirement. 

\begin{figure}[h!tb]
\begin{center}
\includegraphics[width = 0.32 \textwidth]{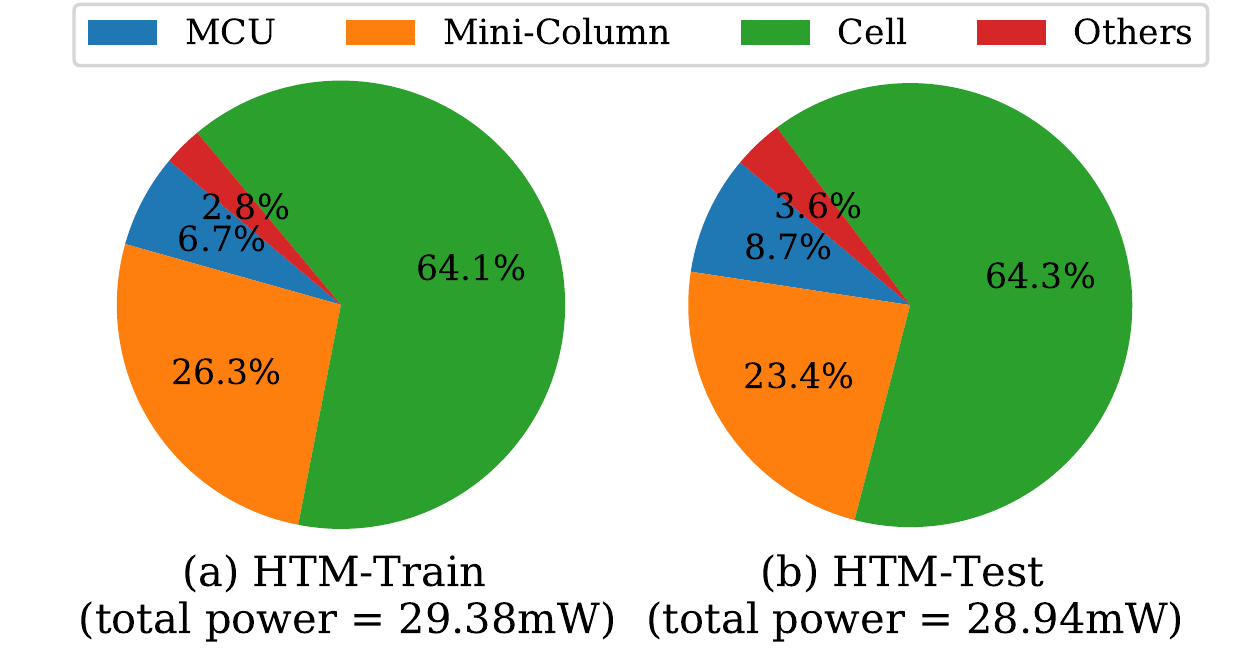}
\caption{The distribution of the power consumption for the building blocks of the proposed HTM system during training and testing modes.}
\label{power_pie}
\end{center}
\end{figure}

\fig{power_pie} shows the distribution of the power consumption among the different entities of the proposed HTM system during the training and testing modes. It implies that in the HTM-Test, most of the power consumption is devoted to the HTM cells as they are more complicated and have a large number of synaptic connections. During the training mode, HTM-Train, the cells and mini-columns pull further power to modulate their synaptic connections. On the contrary, the MCU and other units (arbiter, selector, excluding the H-Tree) consume a small fraction of the total power as they are less complex and have limited memory usage.

In an endeavour to compare our work with previous HTM implementations in literature, we found that performing relative comparisons is a challenging process due to the lack of similarity in network architectures, technology nodes, operating frequency, etc. Thus, we attempt to bring all networks to the same size in terms of the mini-columns and cell count. Also, we hypothesize that the size of the network can be scaled linearly and the same is applied to their power consumption. 
% For the purpose of comparison, we used the power saving formula, given in~\eq{power_eq}
% \begin{equation}
% \label{power_eq}
%     P_{saving} = \frac{P_{ref} - P_{developed}}{P_{ref}} \times 100\% 
% \end{equation}
Starting with Krestinskaya et. al.~\cite{krestinskaya2017hierarchical}\footnote{The authors in this paper also consider linear scaling for the network size and the power consumption.}, here we scaled only the number of mini-columns and the single pixel processing elements (total = 961x1) as detailed information about the distal segments and their sizes are not reported, and this results in 17.45X improvement. In the case of fully CMOS digital design, 77.02X is achieved when compared to our previous work in~\cite{zyarah2019neuromorphic}, and 31.75X and 22.29X when compared to the work done by Li Weifu et al.~\cite{li2019design, li2016hardware}. In contrast to other previous works, the power consumption reported in~\cite{li2016hardware} does not consider the register files, which are usually the most power-hungry components in the design. In the case of~\cite{li2019design}, it is unclear if the register files power consumption is included. It is important to mention here that, in most cases, the overall networks' synaptic connections have not been included in the aforementioned scaling process as there is no clear approach to estimate the power consumption for the individual synaptic primitives. However, since our design uses more synaptic connections, equating our design with previous works in terms of the synaptic connections count may result in further improvement in power consumption. 
\vspace{-2mm}
% =========================================
\section{Conclusions}
This paper proposes a memristor-based mixed-signal architecture of the HTM network including the spatial and temporal aspects of the algorithm. The proposed architecture incorporates several plasticity mechanisms such as synaptogenesis, neurogenesis, etc. that endow the network a high-degree of plasticity with lifelong learning and minimal energy dissipation. The high-level behavioral model of the architecture is verified for time-series data prediction. It is found that the MAPE of the hardware model is more than that in the software counterpart by 1.129X. This degradation is mainly attributed to the memristor devices' non-idealities and the use of synthetic synapses representation. The proposed architecture is also evaluated for latency and lifespan. We found that the mixed-signal implementation is $\approx$3.46X faster than the pure CMOS implementation and its less affected by network scale, while the network elasticity (lifespan) can be up to 8 years, assuming that learning occurs every 10ms. When it comes to network robustness, it is observed that the HTM network is robust to device failure, but this is not the case for its SDR classifier, which is impacted by stuck-on faults. Furthermore, it is observed that the power consumption in the proposed architecture is dominated by the cells, particularly the proximal and distal segments. Thus, in our design, we strive to limit their use to minimum number of cycles and thereby reduce the average total power consumption of the network to 29.38mW.

\begin{IEEEbiography}[{\includegraphics[width=1in,height=1.25in,clip,keepaspectratio]{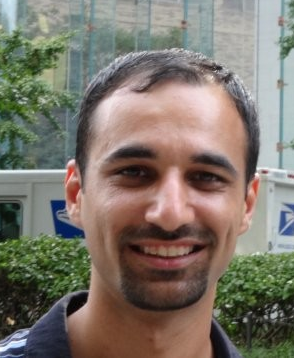}}]{Abdullah M. Zyarah} is a lecturer at the Department of Electrical Engineering, University of Baghdad. He specializes in digital and mixed-signal designs and his current research interests include Neuromorphic architectures for energy constrained platforms and biologically inspired algorithms. Mr. Zyarah received his B.Sc. degree in Electrical Engineering from University of Baghdad, Iraq, in 2009, and M.Sc. degree in the same discipline from Rochester Institute of Technology, USA, in 2015. Currently, he is a PhD candidate within the  Neuromorphic AI Lab research group in the Department of Computer Engineering, Rochester Institute of Technology. 
\end{IEEEbiography}

\vskip -2\baselineskip plus -1fil
\begin{IEEEbiography}[{\includegraphics[width=1in,height=1.25in,clip,keepaspectratio]{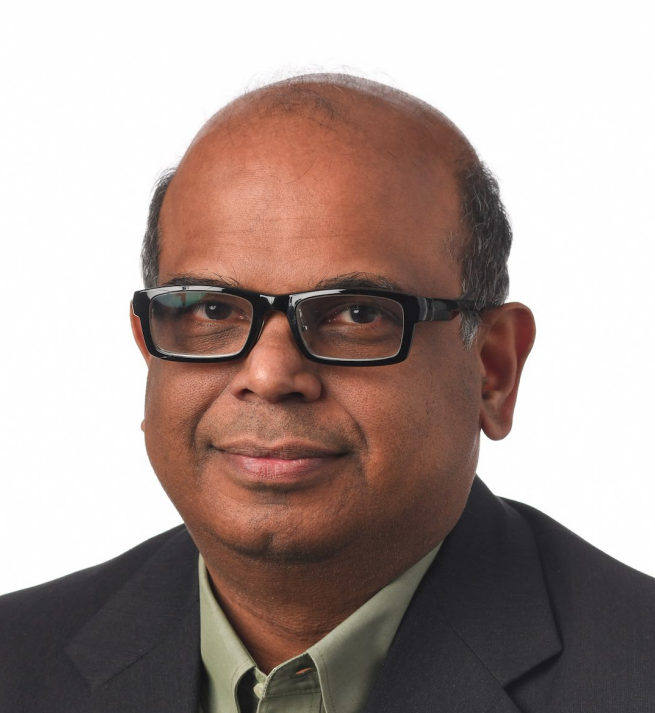}}]{Kevin Gomez} is a Technologist in the Research Group at Seagate. His research interests include computer architecture post Dennard scaling and human brain inspired computing.
\end{IEEEbiography}

\vskip -2\baselineskip plus -1fil
\begin{IEEEbiography}
[{\includegraphics[width=1.25in,height=1.2in,clip,keepaspectratio]{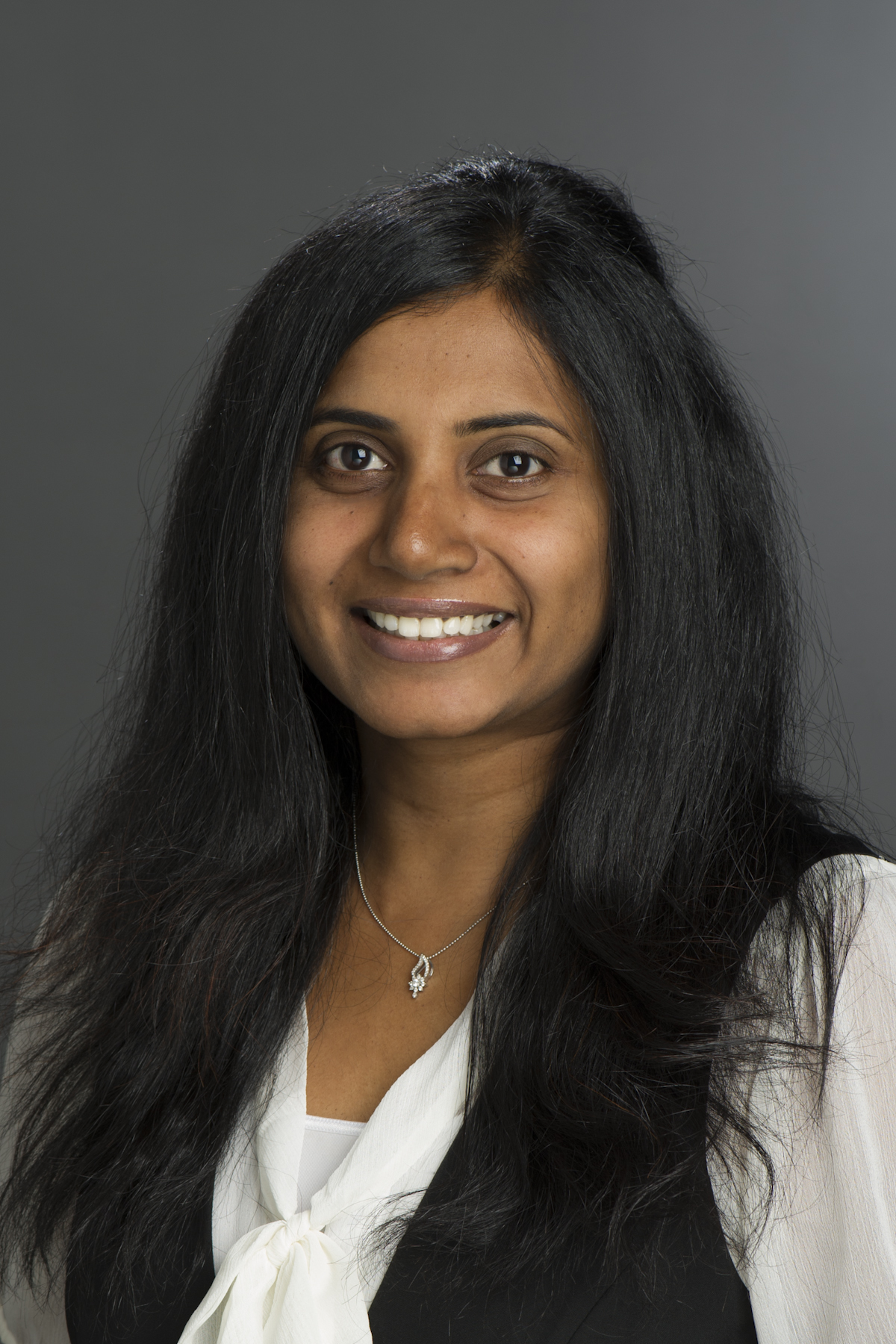}}]{Dr. Dhireesha Kudithipudi}[M'06, SM'16] is a professor and founding Director of the AI Consortium at the University of Texas, San Antonio and Robert F McDermott Chair in Engineering. Her research interests are in neuromorphic AI, low power machine intelligence, brain-inspired accelerators, and use-inspired research. Her team has developed comprehensive neocortex and cerebellum based architectures with nonvolatile memory, hybrid plasticity models, and ultra-low precision architectures.
 
She is passionate about transdisciplinary and inclusive research training in AI fields. She is the recipient of the Clare Booth Luce Scholarship in STEM for women in highered (2018) and the 2018 Technology Women of the Year in Rochester.

\end{IEEEbiography}

\end{document}